\documentclass[dvipdfmx]{article}
\usepackage{latexsym}
\usepackage{amsmath,amssymb}
\usepackage[hiresbb]{graphicx}
\usepackage{tikz}
\usepackage{fancybox,ascmac}
\usepackage{bm}
\usepackage{mathtools,slashed}
\usepackage{fancyhdr}

\begin{document}

\begin{flushright}
DPUR/TH/79\\
March, 2024\\
\end{flushright}
\vspace{20pt}

\pagestyle{empty}
\baselineskip15pt

\begin{center}
{\Large\bf  Conformal Symmetry in Quantum Gravity \vskip 1mm }

\vspace{20mm}

{\large Ichiro Oda\footnote{
           E-mail address:\ ioda@sci.u-ryukyu.ac.jp
                  }

\vspace{10mm}
           Department of Physics, Faculty of Science, University of the 
           Ryukyus,\\
           Nishihara, Okinawa 903-0213, Japan\\          
}

\end{center}


\vspace{10mm}
\begin{abstract}

We study the problem of how to derive conformal symmetry in the framework of quantum gravity.
We start with a generic gravitational theory which is invariant under both the general coordinate transformation (GCT) and 
Weyl transformation (or equivalently, local scale transformation), and then construct its BRST formalism by fixing 
the gauge symmetries by the extended de Donder gauge and scalar gauge conditions. These gauge-fixing conditions are
invariant under global $GL(4)$ and global scale transformations. The gauge-fixed and BRST invariant quantum action possesses 
a huge Poincar\'e-like $IOSp(10|10)$ global symmetry, from which we can construct an extended conformal symmetry 
in a flat Minkowski background in the sense that the Lorentz symmetry is replaced with the $GL(4)$ symmetry. Moreover, 
we construct the conventional conformal symmetry out of this extended symmetry.  With a flat Minkowski background 
$\langle g_{\mu\nu} \rangle = \eta_{\mu\nu}$ and a non-zero scalar field $\langle \phi \rangle \neq 0$, the $GL(4)$ 
and global scale symmetries are spontaneously broken to the Lorentz symmetry, thereby proving that the graviton 
and the dilaton are respectively the corresponding Nambu-Goldstone bosons, and therefore they must be exactly massless 
at nonperturbative level. One of remarkable aspects in our findings is that in quantum gravity, a derivation of conformal symmetry 
does not depend on a classical action, and its generators are built from only the gauge-fixing and the FP ghost actions. 
Finally, we address a generalized Zumino theorem in quantum gravity. 

\end{abstract}

\newpage
\pagestyle{plain}
\pagenumbering{arabic}


\section{Introduction}

There is no question that both local and global symmetries play an important role in elementary-particle physics 
and quantum gravity. For instance, in quantum chromodynamics (QCD)  it has been found that a local symmetry is 
the nonabelian gauge symmetry based on the gauge group $SU(3)$, and that this local symmetry gives rise to 
physically significant effects, such as the asymptotic freedom and the confinement of quarks and gluons. 
In addition to the $SU(3)$ gauge symmetry, there is a $U(1)_V \times U(1)_A$ global symmetry of the quark action. 
The $U(1)_A$ symmetry is anomalous and the effect of a $U(1)_A$ transformation is to change the value of 
the theta angle, which requires us to consider an instanton to make the vacuum energy density to depend on the theta angle.  
 
On the other hand, in quantum gravity, the meaning of symmetries is more subtle than that in elementary-particle physics.
Although general relativity has been beautifully established in the Riemannian geometry on the basis of the general coordinate 
invariance and the equivalence principle, it seems that the feature of the non-renormalizability of general relativity
would need a more huge local symmetry such as a local supersymmetry or a Weyl symmetry (or equivalently, a local scale symmetry).
As for a global symmetry, in classical general relativity, it could be broken by the no-hair theorem of black holes \cite{MTW}.  
Moreover, in string theory, which is a strong candidate of quantum gravity, we never get any additive conservation laws 
and at least in known string vacua, the additive global symmetries turn out to be either gauge symmetries or explicitly 
violated \cite{Banks}.

In such a situation, it might appear to be strange to seek for a new global symmetry in quantum gravity, but 
even in quantum gravity, we have important global symmetries such as the BRST symmetry and the conservation
law of the ghost number. The BRST symmetry, which is a residual global symmetry emerging after the gauge-fixing procedure, 
plays a role in proving the unitarity of the theory and deriving the Ward identities among the Green functions \cite{Kugo-Ojima}. 
On the other hand, the conservation of the ghost number is violated in a two-dimensional quantum theory, 
and leads to the ghost number anomaly, which is closely related to the Riemann-Roch theorem on the closed Riemann 
surfaces \cite{Peskin}.  

From this viewpoint, it is valuable to derive global symmetries, which include the BRST symmetry and the symmetry of 
the FP ghost number, in quantum gravity. If such the symmetries are purely built from quantum fields such as the
FP (anti)ghosts and the Nakanishi-Lautrup auxiliary fields, they could escape from the no-hair theorem of black holes,
and might have some important applications for the study of their anomalies and the confinement of massive ghost
in higher-derivative gravities.      

In a flat Minkowski space-time, conformal symmetry occupies a special position among many global symmetries in that conformal
symmetry is ubiquitous in physics ranging from elementary-particle physics to condensed matter physics \cite{Gross, Nakayama}. 
Even if conformal symmetry 
plays an important role, it must be spontaneously broken in reality since our world has a built-in scale in it. 
Moreover, it is of interest that conformal symmetry is connected with a classical gravity formulated in a curved space-time.
In fact, the Zumino theorem advocates that the theories which are invariant under the GCT and Weyl transformation have 
conformal invariance in the flat Minkowski background at the classical level \cite{Zumino}. Then, it is natural to ask ourselves 
whether the Zumino theorem can be generalized even in quantum gravity or not. In our past works 
\cite{Oda-Q, Oda-W, Oda-Saake, Oda-Ohta}, we have attacked this question. We have shown that in Weyl invariant scalar-tensor 
gravity \cite{Oda-W}, the Zumino theorem is valid whereas in conformal gravity \cite{Oda-Ohta} it is not so.
In these works, since we have specified a classical theory, it is not clear if the answer is independent of the choice
of the classical action or not. It is therefore desirable to start with not a specific action but a more general gravitational action, 
which is invariant under both the GCT and Weyl transformation, and investigate the validity of the Zumino theorem in quantum
gravity. This is one of our motivations behind the present study. Actually, we will find that it is not necessary to fix such a classical 
action at all in order to understand the Zumino theorem in quantum gravity.  

The outline of this article is as follows: In the next section we construct a BRST formalism of our theory. In Section 3, we carry out 
the canonical quantization and calculate the equal-time (anti)commutation relations (ETCRs). 
In Section 4, we prove the existence of a Poincar\'e-like $IOSp(10|10)$ 
global symmetry in a gauge-fixed and BRST invariant quantum Lagrangian and compute its algebra. In Section 5, we derive
conformal symmetry in a flat Minkowski space-time from the Poincar\'e-like $IOSp(10|10)$ symmetry. In Section 6, 
we investigate spontaneous symmetry breakdown of $GL(4)$ symmetry and global scale symmetry to Lorentz
symmetry. Through this mechanism of the symmetry breaking, we can precisely prove that the graviton and the
dilaton are exactly massless owing to the Nambu-Goldstone theorem. The final section is devoted to a conclusion.

Four appendices are put for technical details. In Appendix A, we show that the extended de Donder gauge condition
is invariant under the global $GL(4)$ transformation. In Appendix B, we derive the ETCR, $[ \dot g_{\mu\nu}, b_\rho^\prime ]$.
In Appendix C, we present a proof of the ETCR, 
$[ \dot b_\mu, b_\nu^\prime ] = - i \tilde f \phi^{-2} ( \partial_\mu b_\nu + \partial_\nu b_\mu ) \delta^3$
without recourse to the Einstein's equation. In Appendix D, we give two different proofs for an algebra
$[ P_\mu, K^\nu ] = - 2i ( G^\rho\,_\rho - D ) \delta_\mu^\nu$.

\section{BRST formalism}

We wish to perform a manifestly covariant BRST quantization of a gravitational theory which
is invariant under both general coordinate transformation (GCT) and Weyl transformation (or equivalently,
local scale transformation). To take a more general theory into consideration, without specifying 
a concrete expression of the gravitational Lagrangian, we will start with the following classical 
Lagrangian\footnote{We follow the notation and conventions of MTW textbook \cite{MTW}. Greek 
little letters $\mu, \nu, \cdots$ and Latin ones $i, j, \cdots$ are used for space-time and spatial indices, 
respectively; for instance, $\mu= 0, 1, 2, 3$ and $i = 1, 2, 3$. The Riemann curvature tensor 
and the Ricci tensor are respectively defined by $R^\rho{}_{\sigma\mu\nu} = \partial_\mu \Gamma^\rho_{\sigma\nu} 
- \partial_\nu \Gamma^\rho_{\sigma\mu} + \Gamma^\rho_{\lambda\mu} \Gamma^\lambda_{\sigma\nu} 
- \Gamma^\rho_{\lambda\nu} \Gamma^\lambda_{\sigma\mu}$ and $R_{\mu\nu} = R^\rho{}_{\mu\rho\nu}$. 
The Minkowski metric tensor is denoted by $\eta_{\mu\nu}$; $\eta_{00} = - \eta_{11} = - \eta_{22} 
= - \eta_{33} = -1$ and $\eta_{\mu\nu} = 0$ for $\mu \neq \nu$.} 
\begin{eqnarray}
{\cal L}_c = {\cal L}_c ( g_{\mu\nu}, \phi ),
\label{Lc}  
\end{eqnarray}
which includes the metric tensor field $g_{\mu\nu}$ and a scalar field $\phi$ as dynamical variables.\footnote{It is 
straightforward to add the other fields such as gauge fields and spinor fields.} 
We assume that ${\cal{L}}_c$ does not involve more than first order derivatives of the metric and matter fields.

We are ready to fix the general coordinate symmetry and the Weyl symmetry by suitable 
gauge conditions. It is a familiar fact that after introducing the gauge conditions, instead of such the two local gauge 
symmetries, we are left with two kinds of global symmetries named as the BRST symmetries.  The BRST transformation, 
which is denoted as $\delta_B$, corresponding to the GCT is defined as
\begin{eqnarray}
\delta_B g_{\mu\nu} &=& - ( \nabla_\mu c_\nu+ \nabla_\nu c_\mu)
\nonumber\\
&=& - ( c^\alpha\partial_\alpha g_{\mu\nu} + \partial_\mu c^\alpha g_{\alpha\nu} 
+ \partial_\nu c^\alpha g_{\mu\alpha} ),
\nonumber\\
\delta_B \tilde g^{\mu\nu} &=& \sqrt{-g} ( \nabla^\mu c^\nu+ \nabla^\nu c^\mu 
- g^{\mu\nu} \nabla_\rho c^\rho),
\nonumber\\
\delta_B \phi &=& - c^\lambda \partial_\lambda \phi, \quad 
\delta_B c^\rho = - c^\lambda\partial_\lambda c^\rho, 
\nonumber\\
\delta_B \bar c_\rho &=& i B_\rho, \quad 
\delta_B B_\rho = 0, 
\label{GCT-BRST}  
\end{eqnarray}
where $c^\rho$ and $\bar c_\rho$ are respectively the Faddeev-Popov (FP) ghost and antighost, 
$B_\rho$ is the Nakanishi-Lautrup (NL) field, and we have defined $\tilde g^{\mu\nu} \equiv \sqrt{-g} g^{\mu\nu}$.  
For later convenience, in place of the NL field $B_\rho$ we introduce a new NL field defined as
\begin{eqnarray}
b_\rho= B_\rho- i c^\lambda\partial_\lambda\bar c_\rho,
\label{b-rho-field}  
\end{eqnarray}
and its BRST transformation reads
\begin{eqnarray}
\delta_B b_\rho= - c^\lambda\partial_\lambda b_\rho.
\label{b-BRST}  
\end{eqnarray}
The other BRST transformation, which is denoted as $\bar \delta_B$, 
corresponding to the Weyl transformation is defined as
\begin{eqnarray}
\bar \delta_B g_{\mu\nu} &=& 2 c g_{\mu\nu}, \quad
\bar \delta_B \tilde g^{\mu\nu} = 2 c \tilde g^{\mu\nu},
\nonumber\\
\bar \delta_B \phi &=& - c \phi, \quad 
\bar \delta_B \bar c = i B, \quad 
\bar \delta_B c = \bar \delta_B B = 0, 
\label{Weyl-BRST}  
\end{eqnarray}
where $c$ and $\bar c$ are respectively the FP ghost and FP antighost, and
$B$ is the NL field. Note that the two BRST transformations are nilpotent, i.e.,
\begin{eqnarray}
\delta_B^2 = \bar \delta_B^2 = 0.   
\label{Nilpotent}  
\end{eqnarray}

To complete the two BRST transformations, we have to fix not only the GCT BRST transformation
$\delta_B$ on $c, \bar c$ and $B$ but also the Weyl BRST transformation $\bar \delta_B$ on
$c^\rho, \bar c_\rho$ and $b_\rho$. It is easy to determine the former BRST transformation since the
fields $c, \bar c$ and $B$ are such scalar fields that to match the transformation law for scalar fields
their BRST transformations should take the form:
\begin{eqnarray}
\delta_B B = - c^\lambda\partial_\lambda B, \quad
\delta_B c = - c^\lambda\partial_\lambda c, \quad
\delta_B \bar c = - c^\lambda\partial_\lambda \bar c.   
\label{GCT-BRST2}  
\end{eqnarray}
On the other hand, there is an ambiguity in fixing the latter BRST transformation, but we would like to
propose a recipe for achieving this goal. The recipe \cite{Oda-W} is to just assume that 
the two BRST transformations are anti-commute with each other, that is,   
\begin{eqnarray}
\{ \delta_B, \bar \delta_B \} \equiv \delta_B \bar \delta_B + \bar \delta_B \delta_B = 0,
\label{GCT-Weyl-BRST}  
\end{eqnarray}
which requires us to take 
\begin{eqnarray}
\bar \delta_B b_\rho = \bar \delta_B c^\rho = \bar \delta_B \bar c_\rho = 0.   
\label{Weyl-BRST2}  
\end{eqnarray}

As the gauge condition for the GCT, we will take ``the extended de Donder gauge'' \cite{Oda-W}
\begin{eqnarray}
\partial_\mu ( \tilde g^{\mu\nu} \phi^2 ) = 0,
\label{Ext-de-Donder}  
\end{eqnarray}
which is invariant under the Weyl transformation (\ref{Weyl-BRST}) as required from the condition (\ref{GCT-Weyl-BRST}).
As for the Weyl transformation, we have to choose an appropriate gauge condition, which is invariant under the GCT, that is, 
a scalar quantity. As such a gauge condition, we will take the so-called ``the scalar gauge condition'' \cite{Oda-W}
\begin{eqnarray}
\partial_\mu ( \tilde g^{\mu\nu} \phi \partial_\nu \phi ) = 0,
\label{Scalar-gauge}  
\end{eqnarray}
which can be alternatively written as 
\begin{eqnarray}
\Box \, \phi^2 = 0.
\label{Alt-Scalar-gauge}  
\end{eqnarray}
The key observation here is that both the extended de Donder gauge condition (\ref{Ext-de-Donder}) and the scalar gauge condition 
(\ref{Scalar-gauge}) are invariant under both a ${\it{global}}$ $GL(4)$ transformation and a ${\it{global}}$ scale transformation.\footnote{The unitary 
gauge condition $\phi = \rm{constant}$ and the Lorenz condition $\nabla_\mu S^\mu = 0$ \cite{Oda-Saake}, where $S_\mu$ is a Weyl gauge field,
are also invariant under these global transformations.} For instance, a proof of the $GL(4)$ invariance of the extended Donder gauge condition 
(\ref{Ext-de-Donder}) is given in Appendix A. 

After taking the extended de Donder gauge condition (\ref{Ext-de-Donder}) for the GCT and the scalar gauge condition
(\ref{Scalar-gauge}) for the Weyl transformation, the gauge-fixed and BRST invariant quantum Lagrangian is given by
\begin{eqnarray}
{\cal L}_q &=& {\cal L}_c + {\cal L}_{GF+FP} + \bar {\cal L}_{GF+FP}
\nonumber\\
&=& {\cal L}_c + i \delta_B ( \tilde g^{\mu\nu} \phi^2 \partial_\mu \bar c_\nu )
+ i \bar \delta_B \left[ \bar c \partial_\mu ( \tilde g^{\mu\nu} \phi \partial_\nu \phi ) \right] 
\nonumber\\
&=& {\cal L}_c - \tilde g^{\mu\nu} \phi^2 ( \partial_\mu b_\nu + i \partial_\mu \bar c_\lambda  \partial_\nu c^\lambda )
\nonumber\\
&+& \tilde g^{\mu\nu} \phi \partial_\mu B \partial_\nu \phi - i \tilde g^{\mu\nu} \phi^2 \partial_\mu \bar c 
\partial_\nu c,
\label{q-Lag}  
\end{eqnarray}
where surface terms are dropped. Let us note that this quantum Lagrangian is also invariant under the global $GL(4)$ transformation 
and the global scale transformation. 

Let us rewrite this Lagrangian concisely as
\begin{eqnarray}
{\cal L}_q = {\cal L}_c - \frac{1}{2} \tilde g^{\mu\nu} E_{\mu\nu},
\label{q-Lag2}  
\end{eqnarray}
where we have defined  
\begin{eqnarray}
E_{\mu\nu} \equiv \phi^2 ( \partial_\mu b_\nu + i \partial_\mu \bar c_\lambda  \partial_\nu c^\lambda )
- \phi \partial_\mu B \partial_\nu \phi + i \phi^2 \partial_\mu \bar c \partial_\nu c
+ ( \mu \leftrightarrow \nu ). 
\label{E}  
\end{eqnarray}
Moreover, it is sometimes more convenient to introduce the dilaton $\sigma (x)$ by defining
\begin{eqnarray}
\phi (x) \equiv e^{\sigma (x)},
\label{Dilaton}  
\end{eqnarray}
and rewrite (\ref{q-Lag2}) further into the form  
\begin{eqnarray}
{\cal L}_q = {\cal L}_c - \frac{1}{2} \phi^2 \tilde g^{\mu\nu} \hat E_{\mu\nu}
= {\cal L}_c - \frac{1}{2} e^{2 \sigma (x)} \tilde g^{\mu\nu} \hat E_{\mu\nu},
\label{q-Lag3}  
\end{eqnarray}
where we have defined  
\begin{eqnarray}
\hat E_{\mu\nu} \equiv \partial_\mu b_\nu + i \partial_\mu \bar c_\lambda  \partial_\nu c^\lambda
- \partial_\mu B \partial_\nu \sigma + i \partial_\mu \bar c \partial_\nu c
+ ( \mu \leftrightarrow \nu ). 
\label{hat-E}  
\end{eqnarray}
Note that the relation between $E_{\mu\nu}$ and $\hat E_{\mu\nu}$ is given by
\begin{eqnarray}
E_{\mu\nu} = \phi^2 \hat E_{\mu\nu} = e^{2 \sigma } \hat E_{\mu\nu}.
\label{E vs h-E}  
\end{eqnarray}

The field equations obtained from variations of $b_\nu$, $B$, $c^\rho$, $\bar c_\rho$, $c$ 
and $\bar c$ in the Lagrangian ${\cal L}_q$, take the form:
\begin{eqnarray}
&{}& \partial_\mu ( \tilde g^{\mu\nu} \phi^2 ) = 0, \qquad
\partial_\mu ( \tilde g^{\mu\nu} \phi \partial_\nu \phi ) = 0, 
\nonumber\\
&{}& \partial_\mu ( \tilde g^{\mu\nu} \phi^2 \partial_\nu \bar c_\rho ) = 0, \qquad
\partial_\mu ( \tilde g^{\mu\nu} \phi^2 \partial_\nu c^\rho ) = 0,
\nonumber\\
&{}& \partial_\mu ( \tilde g^{\mu\nu} \phi^2 \partial_\nu\bar c ) = 0, \qquad
\partial_\mu ( \tilde g^{\mu\nu} \phi^2 \partial_\nu c ) = 0.
\label{q-field-eq}  
\end{eqnarray}
Then, the two gauge-fixing conditions in (\ref{q-field-eq}), or equivalently (\ref{Ext-de-Donder}) and 
(\ref{Scalar-gauge}), lead to a very simple d'Alembert-like equation for the dilaton:
\begin{eqnarray}
g^{\mu\nu} \partial_\mu \partial_\nu \sigma = 0.
\label{Dilaton-eq}  
\end{eqnarray}
It is worthwhile to notice that it is not the scalar field $\phi$ but the dilaton $\sigma$ that satisfies 
this type of equation. Furthermore, the extended de Donder gauge condition, i.e., the first field equation 
in (\ref{q-field-eq}) gives us the similar simple equation for the (anti)ghosts:
\begin{eqnarray}
&{}&  g^{\mu\nu} \partial_\mu \partial_\nu \bar c_\rho = 0, \qquad
g^{\mu\nu} \partial_\mu \partial_\nu c^\rho = 0,
\nonumber\\
&{}& g^{\mu\nu} \partial_\mu \partial_\nu\bar c = 0, \qquad
g^{\mu\nu} \partial_\mu \partial_\nu c = 0.
\label{Gh-field-eq}  
\end{eqnarray}

In order to derive the field equation for $B$, it is useful to take the Weyl BRST transformation for the field 
equation for the antighost $\bar c$ and the result again provides a simple equation:
\begin{eqnarray}
g^{\mu\nu} \partial_\mu \partial_\nu B = 0.
\label{B-eq}  
\end{eqnarray}
Finally, the field equation for $b_\rho$ can be obtained by taking the GCT BRST transformation for 
the field equation for $\bar c_\rho$ as follows:\footnote{This field equation has been previously obtained
from the Einstein's equation \cite{N-O-text, Oda-Q, Oda-W}, but this time we use the GCT BRST transformation 
since we have no concrete expression of the Einstein's equation owing to a generic gravitational action.} 
Taking the GCT BRST transformation for the field equation for $\bar c_\rho$, i. e., , i.e., 
$g^{\mu\nu} \partial_\mu \partial_\nu \bar c_\rho = 0$ leads to
\begin{eqnarray}
( - \partial_\lambda g^{\mu\nu} c^\lambda + 2 g^{\mu\lambda} \partial_\lambda c^\nu ) 
\partial_\mu \partial_\nu \bar c_\rho + i g^{\mu\nu} \partial_\mu \partial_\nu B_\rho = 0.
\label{Der-b-eq1}  
\end{eqnarray}
Inserting Eq. (\ref{b-rho-field}) to this equation and arranging terms, we have 
\begin{eqnarray}
g^{\mu\nu} \partial_\mu \partial_\nu b_\rho = - i c^\lambda \partial_\lambda 
(g^{\mu\nu} \partial_\mu \partial_\nu \bar c_\rho) = 0.
\label{b-rho-eq}  
\end{eqnarray}

In other words, setting $X^M = \{ x^\mu, b_\mu, \sigma, B, c^\mu, \bar c_\mu, c, \bar c \}$
\footnote{In the present formalism, it is important to incorporate the mere space-time
coordinates $x^\mu$ into a set of fields $X^M$. This might suggest that $x^\mu$ should be
promoted to a quantum field in a more complete formalism.}, $X^M$ turns out to obey the d'Alembert-like 
equation:
\begin{eqnarray}
g^{\mu\nu} \partial_\mu \partial_\nu X^M = 0.
\label{X-M-eq}  
\end{eqnarray}
This fact, together with the gauge condition $\partial_\mu ( \tilde g^{\mu\nu} \phi^2 ) = 0$,
produces the two kinds of conserved currents
\begin{eqnarray}
{\cal P}^{\mu M} &\equiv& \tilde g^{\mu\nu} \phi^2 \partial_\nu X^M 
= \tilde g^{\mu\nu} \phi^2 \bigl( 1 \overset{\leftrightarrow}{\partial}_\nu X^M \bigr),
\nonumber\\
{\cal M}^{\mu M N} &\equiv& \tilde g^{\mu\nu} \phi^2 \bigl( X^M 
\overset{\leftrightarrow}{\partial}_\nu Y^N \bigr),
\label{Cons-currents}  
\end{eqnarray}
where we have defined $X^M \overset{\leftrightarrow}{\partial}_\mu Y^N \equiv
X^M \partial_\mu Y^N - ( \partial_\mu X^M ) Y^N$.

\section{Canonical quantization and equal-time commutation relations}

In this section, after introducing the canonical (anti)commutation relations (CCRs), we will evaluate various equal-time 
(anti)commutation relations (ETCRs) among fundamental variables. To simplify various expressions, we will obey 
the following abbreviations adopted in the textbook of Nakanishi and Ojima \cite{N-O-text, Nakanishi}
\begin{eqnarray}
[ A, B^\prime ] &=& [ A(x), B(x^\prime) ] |_{x^0 = x^{\prime 0}},
\qquad \delta^3 = \delta(\vec{x} - \vec{x}^\prime), 
\nonumber\\
\tilde f &=& \frac{1}{\tilde g^{00}} = \frac{1}{\sqrt{-g} g^{00}},
\label{abbreviation}  
\end{eqnarray}
where we assume that $\tilde g^{00}$ is invertible. 

Now, in order to carry out the canonical quantization, we perform the integration by parts once
and rewrite the Lagrangian (\ref{q-Lag}) as
\begin{eqnarray}
{\cal L}_q^\prime 
&=& {\cal L}_c + \partial_\mu ( \tilde g^{\mu\nu} \phi^2 ) b_\nu - i \tilde g^{\mu\nu} \phi^2 \partial_\mu \bar c_\lambda  
\partial_\nu c^\lambda + \tilde g^{\mu\nu} \phi \partial_\mu B \partial_\nu \phi 
\nonumber\\
&-& i \tilde g^{\mu\nu} \phi^2 \partial_\mu \bar c \partial_\nu c + \partial_\mu {\cal{V}}^\mu,
\label{Qq-Lag}  
\end{eqnarray}
where a surface term ${\cal{V}}^\mu$ is defined as ${\cal{V}}^\mu \equiv - \tilde g^{\mu\nu} \phi^2 b_\nu$.
Next, let us set up the canonical (anti)commutation relations (CCRs)
\begin{eqnarray}
&{}& [ g_{\mu\nu}, \pi_g^{\rho\lambda\prime} ] = i \frac{1}{2} ( \delta_\mu^\rho\delta_\nu^\lambda 
+ \delta_\mu^\lambda\delta_\nu^\rho) \delta^3,  \quad 
[ \phi, \pi_\phi^\prime ] = i \delta^3,  \quad
[ B, \pi_B^\prime ] = i \delta^3,
\nonumber\\
&{}& \{ c^\sigma, \pi_{c \lambda}^\prime \} = \{ \bar c_\lambda, \pi_{\bar c}^{\sigma\prime} \}
= i \delta_\lambda^\sigma \delta^3,  \quad
\{ c, \pi_c^\prime  \} = \{ \bar c, \pi_{\bar c}^\prime \} = i \delta^3,
\label{CCRs}  
\end{eqnarray}
where the other (anti)commutation relations vanish.
Here the canonical variables are $g_{\mu\nu}, \phi, B, c^\mu, \bar c_\mu, c, \bar c$ and the corresponding canonical
conjugate momenta are $\pi_g^{\mu\nu}, \pi_\phi, \pi_B, \pi_{c \mu}, \pi_{\bar c}^\mu, \pi_c, \pi_{\bar c}$, respectively 
and the $b_\mu$ field is regarded as not a canonical variable but a conjugate momentum of $\tilde g^{0 \mu}$. 

From the Lagrangian (\ref{Qq-Lag}), it is easy to derive canonical conjugate momenta
\begin{eqnarray}
\pi_g^{\mu\nu} &=& \frac{\partial {\cal L}_q^\prime}{\partial \dot g_{\mu\nu}} = \frac{\partial {\cal L}_c}{\partial \dot g_{\mu\nu}}
- \frac{1}{2} \sqrt{-g} \phi^2 ( g^{0\mu} g^{\nu\rho} + g^{0\nu} g^{\mu\rho} - g^{0\rho} g^{\mu\nu} ) b_\rho,  
\nonumber\\
\pi_\phi &=& \frac{\partial {\cal L}_q^\prime}{\partial \dot \phi} = \frac{\partial {\cal L}_c}{\partial \dot \phi} 
+ 2 \tilde g^{0\mu} \phi b_\mu + \tilde g^{0\mu} \phi \partial_\mu B,  \quad
\pi_B = \frac{\partial {\cal L}_q^\prime}{\partial \dot B} = \tilde g^{0 \mu} \phi \partial_\mu \phi, 
\nonumber\\
\pi_{c \sigma} &=& \frac{\partial {\cal L}_q^\prime}{\partial \dot c^\sigma} = - i \tilde g^{0 \mu} \phi^2 \partial_\mu \bar c_\sigma, \quad
\pi_{\bar c}^\sigma = \frac{\partial {\cal L}_q^\prime}{\partial \dot {\bar c}_\sigma} = i \tilde g^{0 \mu} \phi^2 \partial_\mu c^\sigma,
\nonumber\\ 
\pi_c &=& \frac{\partial {\cal L}_q^\prime}{\partial \dot c} = - i \tilde g^{0 \mu} \phi^2 \partial_\mu \bar c,  \quad
\pi_{\bar c} = \frac{\partial {\cal L}_q^\prime}{\partial \dot {\bar c}} = i \tilde g^{0 \mu} \phi^2 \partial_\mu c,
\label{CCM}  
\end{eqnarray}
where we have defined the time derivative such as $\dot g_{\mu\nu} \equiv \frac{\partial g_{\mu\nu}}{\partial t}
\equiv \partial_0 g_{\mu\nu}$, and differentiation of ghosts is taken from the right. At this point, it is worthwhile to point out 
that although we have not specified the classical Lagrangian ${\cal L}_c$ and cannot write down a concrete expression of
the conjugate momentum $\pi_g^{\mu\nu}$, the canonical conjugate momentum $\pi_g^{0\mu}$ generally has the 
following expression
\begin{eqnarray}
\pi_g^{0\mu} = A^\mu + B^{\mu\nu} \partial_\nu \phi+ C^{\mu\nu} b_\nu,
\label{pi(0mu)}  
\end{eqnarray}
where $A^\mu, B^{\mu\nu}$ and $C^{\mu\nu} \equiv - \frac{1}{2} \tilde g^{00} g^{\mu\nu} \phi^2$ 
have no $\dot g_{\mu\nu}$, and $B^{\mu\nu} \partial_\nu \phi$ does not have $\dot \phi$ 
since $\pi_g^{0\mu}$ does not include the dynamics of the metric and the scalar field. 

Henceforth, we will evaluate various nontrivial equal-time (anti)commutation relations (ETCRs) by using 
the canonical (anti)commutation relations (CCRs), field equations and two kinds of BRST transformations. In particular, use
of the GCT and Weyl BRST transformations makes it possible to derive nontrivial ETCRs without appealing to
the Einstein's equation.

Let us first calculate ETCRs, which can be directly derived from the CCRs in Eq. (\ref{CCRs}) 
and the canonical conjugate momenta in Eq. (\ref{CCM}). The CCR, $[ B, \pi_B^\prime ] = i \delta^3$
leads to the ETCR:
\begin{eqnarray}
[ B, \dot \phi^\prime ] = i \tilde f \phi^{-1} \delta^3.
\label{pi(B)-B}  
\end{eqnarray}
The other CCRs, $[ \Phi, \pi_B^\prime ] = 0$ produce
\begin{eqnarray}
[ \Phi, \dot \phi^\prime ] = 0,
\label{pi(B)-Phi}  
\end{eqnarray}
where $\Phi$ denotes a set of canonical variables except $B$, i.e, $\Phi = \{ g_{\mu\nu}, \phi, c^\mu,
\bar c_\mu, c, \bar c \}$. The antiCCRs, $\{ c^\nu, \pi_{c\mu}^\prime \} = \{ \bar c_\mu, \pi_{\bar c}^{\nu\prime} \} 
= i \delta_\mu^\nu \delta^3$ yield
\begin{eqnarray}
\{ c^\nu, \dot{\bar c}_\mu^\prime \} = - \{ \dot c^\nu, \bar c_\mu^\prime \} = - \tilde f \phi^{-2} 
\delta_\mu^\nu \delta^3,
\label{pi(cmu)-cnu}  
\end{eqnarray}
where we have used a useful identity for generic variables $\Phi$ and $\Psi$
\begin{eqnarray}
[ \Phi, \dot \Psi^\prime] = \partial_0 [ \Phi, \Psi^\prime] - [ \dot \Phi, \Psi^\prime],
\label{identity}  
\end{eqnarray}
which holds for the anticommutation relation as well. 
Similarly, the antiCCRs, $\{ c, \pi_c^\prime \} = \{ \bar c, \pi_{\bar c}^\prime \} = i \delta^3$ produce  
\begin{eqnarray}
\{ c, \dot{\bar c}^\prime \} = - \{ \dot c, \bar c^\prime \} = - \tilde f \phi^{-2} \delta^3.
\label{pi(c)-c}  
\end{eqnarray}

Now we are willing to derive a form of the (anti)ETCRs, $[ \Phi, b_\rho^\prime ]$ with $\Phi$ denoting a generic 
field, which are important ETCRs since the Nakanishi-Lautrup field $b_\mu$ in essence generates a translation.
Particularlly, the ETCR, $[ g_{\mu\nu}, b_\rho^\prime ]$ enables us to derive more complicated ETCRs
later, so we present two different derivations, one of which is based on Eq. (\ref{pi(0mu)}) while
the other derivation relies on the GCT BRST transformation.

Let us first work with the CCR:
\begin{eqnarray}
[ \pi_g^{\alpha 0}, g_{\mu\nu}^\prime ] = - i \frac{1}{2} ( \delta_\mu^\alpha \delta_\nu^0 
+ \delta_\mu^0 \delta_\nu^\alpha) \delta^3.
\label{pi(a0)-g}  
\end{eqnarray}
Then, using Eq. (\ref{pi(0mu)}), we find that this CCR produces
\begin{eqnarray}
[ g_{\mu\nu}, b_\rho^\prime ] = - i \tilde f \phi^{-2} ( \delta_\mu^0 g_{\rho\nu} + \delta_\nu^0 g_{\rho\mu} ) \delta^3.
\label{g-b}  
\end{eqnarray}
Incidentally, this ETCR gives us the similar ETCRs:
\begin{eqnarray}
&{}& [ g^{\mu\nu}, b_\rho^\prime ] = i \tilde f \phi^{-2} ( g^{\mu0} \delta_\rho^\nu+ g^{\nu0} \delta_\rho^\mu) \delta^3,
\nonumber\\
&{}& [ \tilde g^{\mu\nu}, b_\rho^\prime ] = i \tilde f \phi^{-2} ( \tilde g^{\mu0} \delta_\rho^\nu+ \tilde g^{\nu0} \delta_\rho^\mu 
- \tilde g^{\mu\nu} \delta_\rho^0 ) \delta^3.
\label{3-g-b}  
\end{eqnarray}
Here we have used the following fact; since a commutator works as a derivation, we can have formulae
\begin{eqnarray}
&{}& [ g^{\mu\nu}, \Phi^\prime ] = - g^{\mu\alpha} g^{\nu\beta} [ g_{\alpha\beta}, \Phi^\prime ],
\nonumber\\
&{}& [ \tilde g^{\mu\nu}, \Phi^\prime ] = - \left( \tilde g^{\mu\alpha} g^{\nu\beta} - \frac{1}{2} \tilde g^{\mu\nu} 
g^{\alpha\beta} \right) [ g_{\alpha\beta}, \Phi^\prime ],
\label{Simple formulae}  
\end{eqnarray}
where $\Phi$ is a generic field. 

As an alternative derivation of Eq. (\ref{g-b}), we make use of the GCT BRST transformation. For this purpose,
let us start with the CCR, $[ g_{\mu\nu}, \bar c_\rho^\prime ] = 0$, and take its GCT BRST transformation:
\begin{eqnarray}
- \{ c^\alpha \partial_\alpha g_{\mu\nu} + \partial_\mu c^\alpha g_{\alpha\nu} 
+ \partial_\nu c^\alpha g_{\alpha\mu}, \bar c_\rho^\prime \} + [ g_{\mu\nu}, i B_\rho^\prime ] = 0.
\label{g-c-BRS}  
\end{eqnarray}
Then, we find that the CCR, $[ g_{\mu\nu}, \pi_{c\rho}^\prime ] = 0$ produces 
\begin{eqnarray}
[ \dot g_{\mu\nu}, \bar c_\rho^\prime ] = 0,
\label{dot-g-bar-c}  
\end{eqnarray}
where Eqs. (\ref{CCM}) and (\ref{identity}) were used. Using this ETCR twice, the relation (\ref{b-rho-field}),
and (\ref{pi(cmu)-cnu}) in Eq. (\ref{g-c-BRS}), we can arrive at our desired ETCR (\ref{g-b}).
Incidentally, the CCRs, $[ g_{\mu\nu}, \pi_{\bar c}^{\rho\prime} ] = [ g_{\mu\nu}, \pi_c^\prime ] 
= [ g_{\mu\nu}, \pi_{\bar c}^\prime ] = 0$ provide the ETCRs between $g_{\mu\nu}$ and FP ghosts:
\begin{eqnarray}
[ g_{\mu\nu}, \dot c^{\rho\prime} ] = [ g_{\mu\nu}, \dot c^\prime ] = [ g_{\mu\nu}, \dot{\bar c}^\prime ] = 0.
\label{g-dot-FP}  
\end{eqnarray}

Moreover, with the help of the CCR, $[ \pi_g^{\alpha 0}, \Phi^\prime ] = 0$, it is easy to derive the remaining
ETCRs with the form of $[ \Phi, b_\rho^\prime ]$ except for $\Phi = b_\mu$, which will be treated later:
\begin{eqnarray}
[ \phi, b_\rho^\prime ] = [ B, b_\rho^\prime ] = [ c^\mu, b_\rho^\prime ] = [ \bar c_\mu, b_\rho^\prime ] 
= [ c, b_\rho^\prime ] = [ \bar c, b_\rho^\prime ] = 0.
\label{Phi-b}  
\end{eqnarray}

Next, let us attempt to derive a form of the (anti)ETCRs, $[ \dot \Phi, b_\rho^\prime ]$ with $\Phi$ denoting 
a generic field. Let us begin by showing the following ETCRs:
\begin{eqnarray}
[ \phi, \dot{\bar c}_\rho^\prime ] = [ \phi, \ddot{\bar c}_\rho^\prime ] = [ \dot \phi, \dot{\bar c}_\rho^\prime ] = 0.
\label{3phi-bar-c}  
\end{eqnarray}
The first equality can be obtained from either Eq. (\ref{pi(B)-Phi}) or the CCR, $[ \phi, \pi_{c\rho}^\prime ] = 0$.
A proof of the second equality needs the field equation for $\bar c_\rho$ in Eq. (\ref{Gh-field-eq}),
from which we have 
\begin{eqnarray}
\ddot{\bar c}_\rho = - \tilde f ( 2 \tilde g^{0i} \partial_i \dot{\bar c}_\rho + \tilde g^{ij} \partial_i \partial_j \bar c_\rho ).
\label{ddot-bar-c}  
\end{eqnarray}
Together with the first equality in (\ref{3phi-bar-c}), this equation makes it possible to show the second
equality, i.e., $[ \phi, \ddot{\bar c}_\rho^\prime ] = 0$. And the last equality in (\ref{3phi-bar-c})
can be obtained from the first and second equalities. Then, taking the GCT BRST transformation of 
the first equality in (\ref{3phi-bar-c}), we have
\begin{eqnarray}
0 &=& \{ - c^\alpha \partial_\alpha \phi, \dot{\bar c}_\rho^\prime \} 
+  [ \phi, i \partial_0 ( b_\rho^\prime + i c^{\alpha\prime} \partial_\alpha \bar c_\rho^\prime ) ] 
\nonumber\\
&=& - \{ c^\alpha, \dot{\bar c}_\rho^\prime \} \partial_\alpha \phi 
+  i [ \phi, \dot b_\rho^\prime ],
\label{G-BRS-phi-bar-c}  
\end{eqnarray}
where Eq. (\ref{3phi-bar-c}) was used again. Using Eqs. (\ref{pi(cmu)-cnu}), (\ref{identity}) and (\ref{Phi-b}),
we are able to obtain
\begin{eqnarray}
[ \dot \phi, b^\prime_\rho ] = - i \tilde f \phi^{-2} \partial_\rho \phi \delta^3.
\label{dot-phi-b-eq} 
\end{eqnarray} 
It turns out that the CCRs, $[ \pi_{c \mu}, \pi_g^{\alpha 0\prime} ] 
= [ \pi^\mu_{\bar c}, \pi_g^{\alpha 0\prime} ] = 0$ give rise to
\begin{eqnarray}
[ \dot{\bar c}_\mu, b^\prime_\rho ] = - i \tilde f \phi^{-2} \partial_\rho \bar c_\mu \delta^3, \qquad
[ \dot c^\mu, b^\prime_\rho ] = - i \tilde f \phi^{-2} \partial_\rho c^\mu \delta^3.
\label{dot-c-b-eq} 
\end{eqnarray} 
Similarly,  the CCRs, $[ \pi_c, \pi_g^{\alpha 0\prime} ] = [ \pi_{\bar c}, \pi_g^{\alpha 0\prime} ] = 0$ give us
\begin{eqnarray}
[ \dot{\bar c}, b^\prime_\rho ] = - i \tilde f \phi^{-2} \partial_\rho \bar c \delta^3, \qquad
[ \dot c, b^\prime_\rho ] = - i \tilde f \phi^{-2} \partial_\rho c \delta^3.
\label{dot-Wc-b-eq} 
\end{eqnarray} 

In order to evaluate $[ \dot B, b^\prime_\rho ]$, we start with the first equality in Eq. (\ref{dot-Wc-b-eq}),
and take its BRST transformation for the Weyl transformation, which immediately leads to the equation:
\begin{eqnarray}
[ \dot B, b^\prime_\rho ] = - i \tilde f \phi^{-2} \partial_\rho B \delta^3.
\label{B-b-CR} 
\end{eqnarray} 
Furthermore, the ETCR, $[ \dot g_{\mu\nu}, b_\rho^\prime ]$ (or equivalently, $[ g_{\mu\nu}, \dot b_\rho^\prime ]$)
is derived in Appendix B, whose result is written out as
\begin{eqnarray}
[ \dot g_{\mu\nu}, b_\rho^\prime ] &=& - i \Bigl\{ \tilde f \phi^{-2} ( \partial_\rho g_{\mu\nu} 
+ \delta_\mu^0 \dot g_{\rho\nu} + \delta_\nu^0 \dot g_{\rho\mu} ) \delta^3 
\nonumber\\
&+& [ ( \delta_\mu^k - 2 \delta_\mu^0 \tilde f \tilde g^{0 k} ) g_{\rho\nu}
+ (\mu\leftrightarrow \nu) ] \partial_k ( \tilde f \phi^{-2} \delta^3 ) \Bigr\}.
\label{dot g-b}  
\end{eqnarray}
Similarly, we can obtain that
\begin{eqnarray}
&{}& [ g_{\mu\nu}, \dot b_\rho^\prime ] = i \Bigl\{ [ \tilde f \phi^{-2} \partial_\rho g_{\mu\nu} 
- \partial_0 ( \tilde f \phi^{-2} ) ( \delta_\mu^0 g_{\rho\nu} + \delta_\nu^0 g_{\rho\mu} ) ] \delta^3 
\nonumber\\
&{}& + [ ( \delta_\mu^k - 2 \delta_\mu^0 \tilde f \tilde g^{0k} ) g_{\rho\nu} + (\mu \leftrightarrow \nu) ]
\partial_k (\tilde f \phi^{-2} \delta^3) \Bigr\},
\nonumber\\
&{}& [ \tilde g^{\mu\nu}, \dot b_\rho^\prime ] = i \Bigl\{ [ \tilde f \phi^{-2} \partial_\rho \tilde g^{\mu\nu} 
+ \partial_0 ( \tilde f \phi^{-2} ) ( \tilde g^{0\mu} \delta_\rho^\nu + \tilde g^{0\nu} \delta_\rho^\mu
- \tilde g^{\mu\nu} \delta_\rho^0 ) ] \delta^3 
\nonumber\\
&{}& - [ ( \tilde g^{\mu k} - 2 \tilde g^{0\mu} \tilde f \tilde g^{0k} ) \delta_\rho^\nu 
- \frac{1}{2} \tilde g^{\mu\nu} ( \delta_\rho^k - 2 \delta_\rho^0 \tilde f \tilde g^{0k} ) 
\nonumber\\
&{}& + (\mu \leftrightarrow \nu) ]
\partial_k (\tilde f \phi^{-2} \delta^3) \Bigr\}.
\label{g-dot b}  
\end{eqnarray}
In particular, the last equality will be useful in computing an algebra of an extended conformal algebra
later. 

At this stage, we would like to prove very important ETCRs:
\begin{eqnarray}
&{}& [ b_\mu, b_\nu^\prime ] = 0,  
\nonumber\\
&{}& [ \dot b_\mu, b_\nu^\prime ] = - i \tilde f \phi^{-2} ( \partial_\mu b_\nu + \partial_\nu b_\mu ) \delta^3. 
\label{b-b}  
\end{eqnarray}
The first equality can be easily shown by taking the GCT BRST transformation of the ETCR, 
$[ b_\mu, \bar c_\nu^\prime ] = 0$. It is a proof of the second equality that we have to rely on the Einstein's
equation \cite{N-O-text}. In the formalism under consideration, we have not specified the classical action, 
so we cannot utilize the Einstein's equation to prove this equality. However, we have found that 
instead of using the Einstein's equation, we can prove the second equality in (\ref{b-b}) by
appealing to the BRST transformation whose proof is given in Appendix C.   

To close this section, we should comment on the ETCRs involving the canonical variable $B$
since the $B$ field is essentially the generator of a global scale transformation. As seen in Eq. (\ref{pi(B)-B}), 
$\dot \phi$ is the canonical conjugate momentum of the $B$ field, and the other ETCRs, which include
$\dot B$, can be calculated from the CCRs and the Weyl BRST transformation. For instance, the CCRs,
$[ B, \pi_{c\mu}^\prime ] = [ B, \pi_{\bar c}^{\mu\prime} ] = [ B, \pi_c^\prime ] = [ B, \pi_{\bar c}^\prime ] = 0$
give us the ETCRs:
\begin{eqnarray}
[ \dot B, c_\mu^\prime ] = [ \dot B, \bar c^{\mu\prime} ] = [ \dot B, c^\prime ] = [ \dot B, \bar c^\prime ] = 0,
\label{dotB-ETCR}  
\end{eqnarray}
where the formula (\ref{identity}) was used. Moreover, taking the Weyl BRST transformation of the last equality
in (\ref{dotB-ETCR}) produces
\begin{eqnarray}
[ \dot B, B^\prime ] = 0.
\label{dotB-B}  
\end{eqnarray}
Similarly, taking the Weyl BRST transformation of $[ \dot{\bar c}, g_{\mu\nu}^\prime ] = 0$, which can be
obtained from the CCR, $[ \pi_c, g_{\mu\nu}^\prime ] = 0$, gives rise to the ETCR:
\begin{eqnarray}
[ \dot B, g_{\mu\nu}^\prime ] = 2 i \tilde f \phi^{-2} g_{\mu\nu} \delta^3.
\label{dotB-g}  
\end{eqnarray}
It is worthwhile to emphasize that via the CCRs, field equations and BRST transformations, we can calculate all the ECTRs 
except for the ones relevant to the metric and its derivatives such as $[ \dot g_{\mu\nu}, g_{\rho\sigma}^\prime ]$ etc.

\section{Poincar\'e-like $IOSp(10|10)$ global symmetry}

As mentioned in Section 2, a set of fields (including the space-time coordinates $x^\mu$)
$X^M \equiv \{ x^\mu, b_\mu, \sigma, B, c^\mu, \bar c_\mu, c, \bar c \}$ obeys a very simple 
d'Alembert-like equation:
\begin{eqnarray}
g^{\mu\nu} \partial_\mu \partial_\nu X^M = 0.
\label{d'Alemb-eq}  
\end{eqnarray}
This equation holds if and only if we adopt the extended de Donder gauge and the scalar
gauge as gauge-fixing conditions for the GCT and the Weyl transformation, respectively. The existence of 
this simple equation suggests that there could be many of conserved currents as defined in 
Eq. (\ref{Cons-currents}).  In this section, we shall show explicitly that there exist such currents 
and consequently we have a Poincar\'e-like $IOSp(10|10)$ global symmetry.

Let us start with the Lagrangian (\ref{q-Lag3}), which can be cast to the form:
\begin{eqnarray}
{\cal L}_q = {\cal L}_c  - \frac{1}{2} \tilde g^{\mu\nu} \phi^2 \hat E_{\mu\nu}.
\label{Choral-Lag}  
\end{eqnarray}
We can further rewrite it into the form
\begin{eqnarray}
{\cal L}_q &=& {\cal L}_c  - \frac{1}{2} \tilde g^{\mu\nu} \phi^2\eta_{NM} \partial_\mu X^M \partial_\nu X^N
\nonumber\\
&=& {\cal L}_c  - \frac{1}{2} \tilde g^{\mu\nu} \phi^2 \partial_\mu X^M \tilde \eta_{MN} \partial_\nu X^N,
\label{Choral-OSp-Lag}  
\end{eqnarray}
where we have introduced an $IOSp(10|10)$ metric $\eta_{NM} = \eta_{MN}^T \equiv \tilde \eta_{MN}$ 
defined as \cite{Kugo}
\begin{eqnarray}
\eta_{NM} = \tilde \eta_{MN} =
\left(
\begin{array}{cc|cc|cc|cc}
     &                \delta_\mu^\nu &     &   &     &  \\ 
\delta^\mu_\nu  &                    &    &    &    &   \\ 
\hline
    &        &                0    &   -1      &     &    &      \\ 
    &        &               -1    &  0       &       &    &    \\
\hline   
    &        &     &    &       &   -i\delta_\mu^\nu  &   & \\  
    &        &     &    &   i\delta^\mu_\nu &   &    & \\
\hline
    &        &                &        &       &     &         &  -i \\  
    &        &                &        &       &     &          i     & \\
\end{array}
\right)_.
\label{OSp-metric}  
\end{eqnarray}
Let us note that this $IOSp(10|10)$ metric $\eta_{NM}$ has the symmetry property such that 
\begin{eqnarray}
\eta_{MN}=(-)^{|M| \cdot |N|} \eta_{NM} = (-)^{|M|} \eta_{NM}=(-)^{|N|} \eta_{NM},
\label{Prop-OSp-metric}  
\end{eqnarray}
where the statistics index $|M|$ is 0 or 1 when $X^M$ is Grassmann-even or 
Grassmann-odd, respectively. This property comes from the fact that $\eta_{MN}$ is `diagonal' 
in the sense that its off-diagonal, Grassmann-even and Grassmann-odd, and vice versa, matrix elements 
vanish, i.e., $\eta_{MN} = 0$ when $|M| \neq |N|$, thereby being $|M| = |N| = |M| \cdot| N|$ in front of 
$\eta_{MN}$ \cite{Kugo}. 

Now we would like to show that the quantum Lagrangian (\ref{Choral-OSp-Lag}) is invariant under 
a Poincar\'e-like $IOSp(10|10)$ global transformation. First, let us focus on the infinitesimal $OSp$ rotation, 
which is defined by
\begin{eqnarray}
\delta X^M = \eta^{ML} \varepsilon_{LN} X^N \equiv \varepsilon^M{}_N X^N,
\label{OSp-rot}  
\end{eqnarray}
where $\eta^{MN}$ is the inverse matrix of $\eta_{MN}$\footnote{Note that $\eta_{MN}$ is a usual
c-number quantity and satisfies the relation, $\eta_{MN} = \eta^{MN}$.}, and the infinitesimal parameter
$\varepsilon_{MN}$ has the following properties:
\begin{eqnarray}
\varepsilon_{MN} = (-)^{1 + |M| \cdot |N|} \varepsilon_{NM}, \qquad
\varepsilon_{MN} X^L = (-)^{|L| (|M| + |N|)} X^L \varepsilon_{MN}.
\label{varepsilon}  
\end{eqnarray}
Moreover, in order to find the conserved current, we assume that the infinitesimal parameter
$\varepsilon_{MN}$ depends on the space-time coordinates $x^\mu$, i.e., 
$\varepsilon_{MN} = \varepsilon_{MN} (x^\mu)$.

Before delving into the invariance of the quantum Lagrangian (\ref{Choral-OSp-Lag}), we should notice that 
under the infinitesimal $OSp$ rotation (\ref{OSp-rot}), the dilaton $\sigma(x)$, 
which is defined as $\phi = e^\sigma$, transforms as
\begin{eqnarray}
\delta \sigma = \eta^{\sigma L} \varepsilon_{LN} X^N = - \varepsilon_{BN} X^N,
\label{Dilaton-OSp}  
\end{eqnarray}
where we have used (\ref{OSp-metric}) and $\eta_{MN} = \eta^{MN}$.
As for the scalar field $\phi(x)$, this transformation for the dilaton amounts to a Weyl transformation
\begin{eqnarray} 
\phi \rightarrow \phi^\prime = e^{\epsilon (x)} \phi, 
\label{Weyl-phi}  
\end{eqnarray}
where the infinitesimal parameter is defined as $\epsilon (x) = - \varepsilon_{BN} X^N$.

To make the classical Lagrangian ${\cal L}_c$ and the compound metric $\tilde g^{\mu\nu} \phi^2$ 
be invariant under the $OSp$ rotation (\ref{OSp-rot}), the transformation (\ref{Weyl-phi})
should accompany a Weyl transformation for the metric tensor field:
\begin{eqnarray} 
g_{\mu\nu} \rightarrow g_{\mu\nu}^\prime = e^{- 2 \epsilon (x)} g_{\mu\nu}. 
\label{Weyl-g}  
\end{eqnarray}
To put it differently, it is necessary to perform the Weyl transformation (\ref{Weyl-g}) 
for the metric when we perform the $OSp$ rotation (\ref{OSp-rot}).

With the proviso that the Weyl transformation (\ref{Weyl-g}) is made, we can prove an invariance 
of the quantum Lagrangian (\ref{Choral-OSp-Lag}) under the $OSp$ rotation (\ref{OSp-rot}), 
which is given by
\begin{eqnarray}
\delta {\cal L}_q = - \tilde g^{\mu\nu} \phi^2 ( \partial_\mu \varepsilon_{NM} X^M \partial_\nu X^N 
+ \varepsilon_{NM} \partial_\mu X^M \partial_\nu X^N ),
\label{Var-OSp-Lag}  
\end{eqnarray}
where we have used the fact that both the classical Lagrangian ${\cal L}_c$ and
the metric $\tilde g^{\mu\nu} \phi^2$ are invariant under the Weyl transformation.
It is easy to prove that the second term on the RHS vanishes owing to the first property in 
Eq. (\ref{varepsilon}). Thus, ${\cal L}_q$ is invariant under the infinitesimal $OSp$ rotation.
The conserved current is then calculated as 
\begin{eqnarray}
\delta {\cal L}_q &=& - \tilde g^{\mu\nu} \phi^2 \partial_\mu \varepsilon_{NM} X^M \partial_\nu X^N
\nonumber\\
&=& - \frac{1}{2} \tilde g^{\mu\nu} \phi^2 \partial_\mu \varepsilon_{NM} \left[ X^M \partial_\nu X^N
- (-)^{|M| \cdot |N|} X^N \partial_\nu X^M \right]
\nonumber\\
&=& - \frac{1}{2} \tilde g^{\mu\nu} \phi^2 \partial_\mu \varepsilon_{NM} \left( X^M \partial_\nu X^N
- \partial_\nu X^M X^N  \right)
\nonumber\\
&=& - \frac{1}{2} \tilde g^{\mu\nu} \phi^2 \partial_\mu \varepsilon_{NM} 
X^M \overset{\leftrightarrow}{\partial}_\nu X^N
\nonumber\\
&\equiv& - \frac{1}{2} \partial_\mu \varepsilon_{NM} {\cal M}^{\mu MN},
\label{OSp-current}  
\end{eqnarray}
from which the conserved current ${\cal M}^{\mu MN}$ for the $OSp$ rotation takes the form: 
\begin{eqnarray}
{\cal M}^{\mu MN} = \tilde g^{\mu\nu} \phi^2 X^M \overset{\leftrightarrow}{\partial}_\nu X^N.
\label{OSp-current-M}  
\end{eqnarray}

In a similar way, we can derive the conserved current for the infinitesimal translation
\begin{eqnarray}
\delta X^M = \varepsilon^M,
\label{transl}  
\end{eqnarray}
where $\varepsilon^M$ is the infinitesimal parameter and assume that it is a local one
for deriving the corresponding conserved current. Indeed, we can show that ${\cal L}_q$ is invariant 
under the infinitesimal translation as follows
\begin{eqnarray}
\delta {\cal L}_q &=& - \tilde g^{\mu\nu} \phi^2 \eta_{NM} \partial_\mu \varepsilon^M \partial_\nu X^N
\nonumber\\
&=& - \tilde g^{\mu\nu} \phi^2 \partial_\mu \varepsilon_N \partial_\nu X^N
\nonumber\\
&\equiv& - \partial_\mu \varepsilon_M {\cal P}^{\mu M},
\label{transl-current}  
\end{eqnarray}
which implies that the conserved current ${\cal P}^{\mu M}$ for the translation reads 
\begin{eqnarray}
{\cal P}^{\mu M} = \tilde g^{\mu\nu} \phi^2 \partial_\nu X^M
= \tilde g^{\mu\nu} \phi^2 \left( 1 \overset{\leftrightarrow}{\partial}_\nu X^M \right).
\label{transl-current-P}  
\end{eqnarray}
In this case as well, we need to perform an appropriate Weyl transformation for
the metric since the dilaton transforms as $\delta \sigma = \varepsilon^\sigma (x)$ in Eq. (\ref{transl}). 

An important remark is relevant to the expression of the conserved currents (\ref{OSp-current-M}) 
and (\ref{transl-current-P}). To make the quantum Lagrangian ${\cal L}_q$ be invariant 
under the $IOSp(10|10)$ transformation, it is necessary to perform a Weyl transformation for the metric. 
Then, there could be a possibility that the expression of the currents 
might be modified because of this associated Weyl transformation. The good news is that  
as shown in Refs. \cite{Jackiw, Campigotto, Oda-U, Oda-C, Alonso, Oda-V}, the current for the Weyl transformation 
identically vanishes since the Weyl transformation does not involve the derivative of the transformation
parameter as in $\delta g_{\mu\nu} = 2 \Lambda g_{\mu\nu}$ unlike the conventional gauge transformation
as in $\delta A_\mu = \partial_\mu \Lambda$. Thus, although we perform the Weyl transformation, the conserved 
currents (\ref{OSp-current-M}) and (\ref{transl-current-P}) remain unchanged.  
From the conserved currents (\ref{OSp-current-M}) and (\ref{transl-current-P}), the corresponding
conserved charges become
\begin{eqnarray}
M^{MN} &\equiv& \int d^3 x \, {\cal M}^{0 MN} = \int d^3 x \, \tilde g^{0 \nu} \phi^2  
X^M \overset{\leftrightarrow}{\partial}_\nu X^N,
\nonumber\\
P^M &\equiv& \int d^3 x \, {\cal P}^{0 M} = \int d^3 x \, \tilde g^{0 \nu} \phi^2 \partial_\nu X^M.
\label{IOSp-charge}  
\end{eqnarray}
It then turns out that using various ETCRs obtained so far, the $IOSp(10|10)$ generators $\{ M^{MN}, P^M \}$ 
generate an $IOSp(10|10)$ algebra
\begin{eqnarray}
&{}& [ P^M, P^N \} = 0, 
\nonumber\\
&{}& [ M^{MN}, P^R \} = i \bigl[ P^M \tilde \eta^{NR} - (-)^{|N| |R|} P^N \tilde \eta^{MR} \bigr],
\nonumber\\
&{}& [ M^{MN}, M^{RS} \} = i \bigl[ M^{MS} \tilde \eta^{NR} - (-)^{|N| |R|} M^{MR} \tilde \eta^{NS} 
- (-)^{|N| |R|} M^{NS} \tilde \eta^{MR} 
\nonumber\\
&{}& + (-)^{|M| |R| + |N| |S|} M^{NR} \tilde \eta^{MS} \bigr],
\label{IOSp-algebra}  
\end{eqnarray}
where the graded bracket is defined as $[ A, B \} = A B - (-)^{|A| |B|} B A$.

Actually, it is a little tedious to derive the $IOSp(10|10)$ algebra (\ref{IOSp-algebra}) from the ETCRs in a direct manner, 
so we present a different and more concise derivation here. The key observation is that as mentioned in the proof 
of the $IOSp(10|10)$ symmetry of the quantum Lagrangian ${\cal{L}}_q$, in order to make the classical Lagrangian 
${\cal{L}}_c$ and the composite metric $\tilde g^{\mu\nu} \phi^2$ be invariant under the $IOSp(10|10)$ transformation,
we simultaneously need to perform an associated Weyl transformation for the metric. 
But, then this Weyl transformation makes no effect on the conserved currents.  
This fact implies that as long as the $IOSp(10|10)$ charges and their algebra are concerned, one does not have to 
care about such Weyl invariant quantities and can ignore any contribution from them.   

Under such a situation, one can regard $X^M$ in the Lagrangian (\ref{Choral-OSp-Lag}) as only a canonical variable 
and introduce its corresponding canonical conjugate momentum $\pi_M$ as 
\begin{eqnarray}
\pi_M \equiv \frac{\partial {\cal{L}}_q}{\partial \dot X^M} = - \tilde g^{0\mu} \phi^2 \eta_{MN} 
\partial_\mu X^N,
\label{Pi-X}  
\end{eqnarray} 
where the differentiation is taken from the right. The graded CCR, $[ X^M, \pi_N^\prime \} = i \delta_N^M \delta^3$
leads to
\begin{eqnarray}
[ X^M, \dot X^{N\prime} \} = - [ \dot X^M, X^{N\prime} \} = - i \tilde f \phi^{-2} \tilde \eta^{MN} \delta^3.
\label{CCR-X}  
\end{eqnarray} 
From this CCR, it is easy to obtain the following algebra:
\begin{eqnarray}
&{}& [ M^{MN}, X^R \} =  i \left( X^M \tilde \eta^{NR} - (-)^{|N||R|} X^N \tilde \eta^{MR} \right),
\nonumber\\
&{}& [ M^{MN}, \partial_\nu X^R \} =  i \left( \partial_\nu X^M \tilde \eta^{NR} - (-)^{|N||R|} \partial_\nu X^N \tilde \eta^{MR} \right),
\nonumber\\
&{}& [ P^M, X^R \} =  i \tilde \eta^{MR}, \qquad
[ P^M, \partial_\nu X^R \} = 0.
\label{alg-X}  
\end{eqnarray} 

Then, on the basis of this algebra, it is straightforward to show the $IOSp(10|10)$ algebra (\ref{IOSp-algebra}).
For instance, $[ M^{MN}, M^{RS} \}$ can be calculated as follows:
\begin{eqnarray}
&{}& [ M^{MN}, M^{RS} \} = [ M^{MN}, \int d^3 x \, \tilde g^{0\nu} \phi^2 X^R \overset{\leftrightarrow}{\partial}_\nu X^S \} 
\nonumber\\
&{}& = \int d^3 x \, \tilde g^{0\nu} \phi^2 [ M^{MN}, X^R \partial_\nu X^S - \partial_\nu X^R X^S \}
\nonumber\\
&{}& = \int d^3 x \, \tilde g^{0\nu} \phi^2 \Bigl( [ M^{MN}, X^R \} \partial_\nu X^S + (-)^{(|M| + |N|) |R|} X^R 
[ M^{MN}, \partial_\nu X^S \} 
\nonumber\\
&{}& - [ M^{MN}, \partial_\nu X^R \} X^S - (-)^{(|M| + |N|) |R|} \partial_\nu X^R [ M^{MN}, X^S \}  \Bigr)
\nonumber\\
&{}& = i \int d^3 x \, \tilde g^{0\nu} \phi^2 \Bigl[ ( X^M \tilde \eta^{NR} - (-)^{|N||R|} X^N \tilde \eta^{MR} ) \partial_\nu X^S
\nonumber\\
&{}& + (-)^{(|M| + |N|) |R|} X^R ( \partial_\nu X^M \tilde \eta^{NS} - (-)^{|N||S|} \partial_\nu X^N \tilde \eta^{MS} ) 
\nonumber\\
&{}& - ( \partial_\nu X^M \tilde \eta^{NR} - (-)^{|N||R|} \partial_\nu X^N \tilde \eta^{MR} ) X^S
\nonumber\\
&{}& - (-)^{(|M| + |N|) |R|} \partial_\nu X^R ( X^M \tilde \eta^{NS} - (-)^{|N||S|} X^N \tilde \eta^{MS} ) \Bigr]
\nonumber\\
&{}& = i \int d^3 x \, \tilde g^{0\nu} \phi^2 \Bigl[ X^M \overset{\leftrightarrow}{\partial}_\nu X^S \tilde \eta^{NR} 
+ (-)^{(|M| + |N|) |R|} X^R \overset{\leftrightarrow}{\partial}_\nu X^M \tilde \eta^{NS} 
\nonumber\\
&{}& - (-)^{|N||R|} X^N \overset{\leftrightarrow}{\partial}_\nu X^S \tilde \eta^{MR} 
- (-)^{(|M| + |N|) |R| + |N||S|} X^R \overset{\leftrightarrow}{\partial}_\nu X^N \tilde \eta^{MS} \Bigr]
\nonumber\\
&{}& = i \bigl[ M^{MS} \tilde \eta^{NR} + (-)^{( |M| + |N| ) |R|} M^{RM} \tilde \eta^{NS} 
- (-)^{|N| |R|} M^{NS} \tilde \eta^{MR} 
\nonumber\\
&{}& - (-)^{(|M| + |N|) |R| + |N| |S|} M^{RN} \tilde \eta^{MS} \bigr].
\nonumber\\
&{}& = i \bigl[ M^{MS} \tilde \eta^{NR} - (-)^{|N| |R|} M^{MR} \tilde \eta^{NS} 
- (-)^{|N| |R|} M^{NS} \tilde \eta^{MR} 
\nonumber\\
&{}& + (-)^{|M| |R| + |N| |S|} M^{NR} \tilde \eta^{MS} \bigr].
\label{MM-algebra}  
\end{eqnarray}
At the last equality, we have used the relation 
\begin{eqnarray}
M^{MN} =  - (-)^{|M| |N|} M^{NM}.
\label{anti-MM}  
\end{eqnarray}

As a final remark, it is worthwhile to point out that all the global symmetries existing in the 
present theory are expressed in terms of the generators of the Poincar\'e-like $IOSp(10|10)$ 
global symmetry. For instance, the BRST charges for the GCT and Weyl transformation are respectively 
described as
\begin{eqnarray}
Q_B &\equiv& M (b_\rho, c^\rho) = \int d^3 x \, \tilde g^{0 \nu} \phi^2 
b_\rho \overset{\leftrightarrow}{\partial}_\nu c^\rho,  
\nonumber\\
\bar Q_B &\equiv& M (B, c) = \int d^3 x \, \tilde g^{0 \nu} \phi^2 
B \overset{\leftrightarrow}{\partial}_\nu c.
\label{Choral-Symm}  
\end{eqnarray}

\section{Conformal symmetry}

In this section, we wish to explain that the Poincar\'e-like $IOSp(10|10)$ global symmetry, which was established
in the previous section, includes conformal symmetry as a subgroup when the background is fixed to be 
a flat Minkowski space-time.

Before doing that, let us review conformal symmetry briefly. 
Conformal transformation can be defined as the general coordinate transformation which can be undone 
by a Weyl transformation when the space-time metric is the flat Minkowski one \cite{Gross, Nakayama}. 
With this definition, the conformal transformation is described by the equation
\begin{eqnarray}
\partial_\mu \epsilon_\nu + \partial_\nu \epsilon_\mu = 2 \Lambda(x) \eta_{\mu\nu},
\label{Conf-Killing}  
\end{eqnarray} 
where $\epsilon_\mu (x)$ and $\Lambda(x)$ are the infinitesimal transformation parameters of the GCT and
the Weyl transformation, respectively. Taking the trace of Eq. (\ref{Conf-Killing}) enables us to determine $\Lambda(x)$ to be
\begin{eqnarray}
\Lambda = \frac{1}{4} \partial^\rho \epsilon_\rho.
\label{lambda-0}  
\end{eqnarray} 
Inserting this $\Lambda$ to Eq. (\ref{Conf-Killing}) yields 
\begin{eqnarray}
\partial_\mu \epsilon_\nu + \partial_\nu \epsilon_\mu = \frac{1}{2} \partial^\rho \epsilon_\rho \eta_{\mu\nu},
\label{Conf-Killing2}  
\end{eqnarray} 
which is often called the ``conformal Killing equation'' in the Minkowski space-time. 
It is well-known that a general solution to the conformal Killing equation reads
\begin{eqnarray}
\epsilon^\mu = a^\mu + \omega^{\mu\nu} x_\nu + \lambda x^\mu + k^\mu x^2 - 2 x^\mu k_\rho x^\rho,
\label{Conf-Killing-vector}  
\end{eqnarray} 
where $a^\mu, \omega^{\mu\nu} = - \omega^{\nu\mu}, \lambda$ and $k^\mu$ are all constant parameters and 
they correspond to the translation, the Lorentz transformation, the dilatation and the special conformal 
transformation, respectively. Substituting Eq. (\ref{Conf-Killing-vector}) into Eq. (\ref{lambda-0}) gives us 
\begin{eqnarray}
\Lambda = \lambda - 2 k_\mu x^\mu.
\label{lambda-2}  
\end{eqnarray} 

We are now ready to clarify the presence of conformal symmetry in the formalism at hand.
To do so, let us note that although we have already fixed general coordinate invariance and Weyl one 
by the extended de Donder gauge condition (\ref{Ext-de-Donder}) and the scalar gauge one (\ref{Scalar-gauge}),
respectively, we are still left with its linearized, residual symmetries. 
In order to look for such residual symmetries, it is convenient to begin by the extended de Donder gauge (\ref{Ext-de-Donder}) 
and take its variation under the GCT.\footnote{A similar strategy has been adopted in different theories in Refs. 
\cite{Ferrari, Oda-R, Kamimura, Oda-RWS}.}  As seen in Eq. (\ref{GCT-BRST}), the GCT is defined as
\begin{eqnarray}
\delta g_{\mu\nu} &=& - ( \varepsilon^\alpha\partial_\alpha g_{\mu\nu} + \partial_\mu \varepsilon^\alpha g_{\alpha\nu} 
+ \partial_\nu \varepsilon^\alpha g_{\mu\alpha} ),
\nonumber\\
\delta \tilde g^{\mu\nu} &=& - \partial_\alpha ( \varepsilon^\alpha \tilde g^{\mu\nu} ) 
+ \partial_\alpha \varepsilon^\mu \tilde g^{\alpha\nu} + \partial_\alpha \varepsilon^\nu \tilde g^{\mu\alpha},
\nonumber\\
\delta \phi &=& - \varepsilon^\alpha \partial_\alpha \phi, 
\label{GCT}  
\end{eqnarray}
where $\varepsilon^\alpha$ is an infinitesimal transformation parameter since replacing $\varepsilon^\alpha$ with
the ghost $c^\alpha$ produces the GCT BRST in (\ref{GCT-BRST}). By using Eq. (\ref{GCT})
and the extended de Donder gauge condition (\ref{Ext-de-Donder}), the variation of the extended de Donder gauge 
condition (\ref{Ext-de-Donder}) takes the form
\begin{eqnarray}
\delta \left[ \partial_\mu ( \tilde g^{\mu\nu} \phi^2 ) \right] = \partial_\mu \partial_\alpha \varepsilon^\nu 
\tilde g^{\mu\alpha} \phi^2.
\label{Var-Ext-de-Donder}  
\end{eqnarray}
Thus, a residual symmetry for the GCT exists when the following equation is obeyed:
\begin{eqnarray}
\partial_\mu \partial_\alpha \varepsilon^\nu = 0.
\label{Res-GCT}  
\end{eqnarray}
It is obvious that with a flat Minkowski metric $g_{\mu\nu} = \eta_{\mu\nu}$, there is a zero-mode solution 
to Eq. (\ref{Res-GCT}) which is linear in $x^\mu$
\begin{eqnarray}
\varepsilon^\mu = A^\mu{}_\nu x^\nu + a^\mu,
\label{GC4-transl}  
\end{eqnarray}
where $A^\mu{}_\nu$ and $a^\mu$ are constant quantities and here indices are raised and lowered
with the flat Minkowski metric $\eta_{\mu\nu}$. This equation implies that we have a global $GL(4)$ 
symmetry and a translation symmetry associated with $A^\mu{}_\nu$ and $a^\mu$, respectively, as residual symmetries.
To put it differently, the extended de Donder gauge condition (\ref{Ext-de-Donder}) leaves invariances
under the GCT with the transformation parameter $\varepsilon^\mu (x)$ linear in the coordinates $x^\mu$,
which precisely correspond to the global $GL(4)$ and translational invariances. 

Next, along the similar line of arguments, let us start with the scalar gauge condition (\ref{Scalar-gauge})
and take its variation under Weyl transformation, which is defined as
\begin{eqnarray}
\bar \delta g_{\mu\nu} &=& 2 \Lambda g_{\mu\nu}, \qquad
\bar \delta \tilde g^{\mu\nu} = 2 \Lambda \tilde g^{\mu\nu}, 
\nonumber\\
\bar \delta \phi &=& - \Lambda \phi,  \qquad
\bar \delta \sigma = - \Lambda,
\label{Weyl-transf}  
\end{eqnarray}
where $\Lambda$ is an infinitesimal transformation parameter. Note that the replacement of
$\Lambda$ with the ghost $c$ in Eq. (\ref{Weyl-transf}) leads to  the Weyl BRST transformation in 
Eq. (\ref{Weyl-BRST}) as required.
The result reads
\begin{eqnarray}
\bar \delta \left[ \partial_\mu ( \tilde g^{\mu\nu} \phi \partial_\nu \phi ) \right] 
= \bar \delta \left[ \partial_\mu ( \tilde g^{\mu\nu} \phi^2 \partial_\nu \sigma ) \right] 
= - \tilde g^{\mu\nu} \phi^2 \partial_\mu \partial_\nu \Lambda,
\label{Var-Scalar-Gauge}  
\end{eqnarray}
where the extended de Donder gauge condition (\ref{Ext-de-Donder}) was used.
We therefore find that a residual symmetry for the Weyl transformation exists when the equation
\begin{eqnarray}
g^{\mu\nu} \partial_\mu \partial_\nu \Lambda = 0,
\label{Res-Weyl}  
\end{eqnarray}
is satisfied. With the flat Minkowski metric $g_{\mu\nu} = \eta_{\mu\nu}$, this equation has a linearized
zero-mode solution 
\begin{eqnarray}
\Lambda = \lambda - 2 k_\mu x^\mu,
\label{Scale-Special-conf}  
\end{eqnarray}
where $\lambda$ and $k_\mu$ are constants. Here the coefficient $-2$ in front of the last term is
chosen for convenience. Comparing (\ref{Scale-Special-conf}) with (\ref{lambda-2}), we find that 
the transformations associated with the parameters $\lambda$ and $k_\mu$, respectively, correspond to 
dilatation and special conformal transformation in a flat Minkowski background.\footnote{For clarity, 
we will call a global scale transformation in a flat Minkowski space-time ``dilatation''. Dilatation is usually 
interpreted as a subgroup of the general coordinate transformation in a such way that 
the space-time coordinates are transformed as $x^\mu \rightarrow \Omega x^\mu$ 
in the flat space-time where $\Omega$ is a constant scale factor, whereas the global scale 
transformation is a rescaling of all lengths by the same $\Omega$ by 
$g_{\mu\nu} \rightarrow \Omega^2 g_{\mu\nu}$. The two viewpoints are completely equivalent 
since all the lengths are defined via the line element $d s^2 = g_{\mu\nu} d x^\mu d x^\nu$.}   
In other words, in this case, finding the residual symmetries is equivalent to solving the conformal Killing equation.

We can also verify the invariance of the quantum Lagrangian under the residual symmetries more directly.  
As an illustration, let us take the residual symmetries (\ref{Scale-Special-conf}) stemmed from the Weyl transformation
into consideration. It is easy to check that the quantum Lagrangian (\ref{q-Lag3}) is invariant under the residual 
symmetries
\begin{eqnarray}
&{}& \delta g_{\mu\nu} = 2 ( \lambda - 2 k_\rho x^\rho ) g_{\mu\nu},
\nonumber\\
&{}& \delta \sigma = - ( \lambda - 2 k_\rho x^\rho ), \qquad
\delta b_\mu = 2 k_\mu B,
\label{Res-symm}  
\end{eqnarray}
where the other fields are unchanged. Then, generators corresponding to the transformation parameters
$\lambda$ and $k_\mu$ are respectively constructed out of those of the Poincar\'e-like $IOSp(10|10)$ symmetry as
\begin{eqnarray}
D_0 &\equiv& - P^B \equiv - P(B) = - \int d^3 x \, \tilde g^{0 \nu} \phi^2 \partial_\nu B,
\nonumber\\
K^\mu &\equiv& 2 M^{x^\mu B} \equiv 2 M^\mu (x, B) = 2 \int d^3 x \, \tilde g^{0 \nu} \phi^2 x^\mu
\overset{\leftrightarrow}{\partial}_\nu B.
\label{Res-gen}  
\end{eqnarray}
For instance, we can explicitly show that these generators generate the former transformation in Eq. (\ref{Res-symm})
as follows
\begin{eqnarray}
\delta g_{\mu\nu} (x) &=& [ i ( \lambda D_0 + k_\rho K^\rho ),  g_{\mu\nu} (x) ]
\nonumber\\
&=& i \int d^3 y [ \tilde g^{0 \sigma} \phi^2 (y) ( - \lambda \partial_\sigma B (y)
+ 2 k_\rho y^\rho \overset{\leftrightarrow}{\partial}_\sigma B (y) ), g_{\mu\nu} (x) ]
\nonumber\\
&=& i \int d^3 y \, \tilde g^{00} \phi^2 (y) ( - \lambda + 2 k_\rho y^\rho ) [ \dot B (y), g_{\mu\nu} (x) ]
\nonumber\\
&=& 2 ( \lambda - 2 k_\rho x^\rho ) g_{\mu\nu} (x),
\label{Check-gen}  
\end{eqnarray}
where the ETCR (\ref{dotB-g}) was used in the last equality.

In addition to the generators $D_0$ and $K^\mu$, one can also construct the translation generator $P_\mu$
and $GL(4)$ generator $G^\mu{}_\nu$ from those of the Poincar\'e-like $IOSp(10|10)$ symmetry as
\begin{eqnarray}
P_\mu &\equiv& P_{b_\mu} \equiv P_\mu (b) = \int d^3 x \, \tilde g^{0 \nu} \phi^2 \partial_\nu b_\mu,
\nonumber\\
G^\mu{}_\nu &\equiv& M^{x^\mu}{}_{b_\nu} - i M^{c^\mu}{}_{\bar c_\nu} \equiv M^\mu{}_\nu (x, b) - i M^\mu{}_\nu (c^\tau, \bar c_\lambda)
\nonumber\\
&=& \int d^3 x \, \tilde g^{0 \lambda} \phi^2 ( x^\mu \overset{\leftrightarrow}{\partial}_\lambda b_\nu
- i c^\mu \overset{\leftrightarrow}{\partial}_\lambda \bar c_\nu ).
\label{Trans-GL}  
\end{eqnarray}

At this point, let us elaborate a bit on the special conformal generator $K^\mu$. It is well-known that
under the special conformal transformation an arbitrary ``{\it{primary}}'' operators ${\cal{O}}_i (x)$ of conformal dimension 
$\Delta_i$ transforms as \cite{Gross, Nakayama}
\begin{eqnarray}
[ i K_\mu, {\cal{O}}_i (x) ] = ( 2 x_\mu x^\nu \partial_\nu - x^2 \partial_\mu ) {\cal{O}}_i (x)
+ 2 \Delta_i x_\mu {\cal{O}}_i (x) - 2 x^\nu ( S_{\mu\nu} )_i{}^j {\cal{O}}_j (x),
\label{Spec-Conf}  
\end{eqnarray}
where $( S_{\mu\nu} )_i{}^j$ is a spin matrix for the Lorentz transformation.  In our formalism, it is easy to show that 
under the special conformal transformation the fundamental variables $\Phi \equiv \{ g_{\mu\nu}, \phi, b_\mu, 
B, c_\mu, \bar c^\mu, c, \bar c\}$ transform as
\begin{eqnarray}
&{}& [ i K^\rho, g_{\mu\nu} ] = - 4 x^\rho g_{\mu\nu},  \qquad
[ i K^\rho, \phi ] = 2 x^\rho \phi,  \qquad
[ i K^\rho, b_\mu ] = 2 \delta_\mu^\rho B,  \qquad
\nonumber\\
&{}& [ i K^\rho, B ] = [ i K^\rho, c_\mu ] = [ i K^\rho, \bar c^\mu ] 
= [ i K^\rho, c ] = [ i K^\rho, \bar c ] = 0.
\label{Spec-Conf-us}  
\end{eqnarray}
Comparing Eq. (\ref{Spec-Conf-us}) to Eq. (\ref{Spec-Conf}), we can immediately notice three things: 
Firstly, $g_{\mu\nu}$ and $\phi$ are primary fields of conformal dimension $-2$ and $1$, respectively, and 
$B, c_\mu, \bar c^\mu, c$ and $\bar c$ are all primary fields of conformal dimension $0$ 
whereas $b_\mu$ is not a primary field.  Secondly, there is no term involving the terms quadratic in $x^\mu$
in Eq. (\ref{Spec-Conf-us}) since we have confined ourselves to linearized symmetries in $x^\mu$. Finally, 
there is not the term including the spin matrix $( S_{\mu\nu} )_i{}^j$ in Eq. (\ref{Spec-Conf-us}) since we do not 
consider spinor fields. 

Now we would like to show that in our theory there is an extended conformal algebra in a flat Minkowski space-time 
where the Lorentz symmetry is replaced with the $GL(4)$ symmetry. For this aim, let us consider 
a set of generators, $\{ P_\mu, G^\mu{}_\nu, K^\mu, D_0 \}$, which has 25 independent generators and
consists of a larger symmetry group than conformal symmetry constructed in terms of 15 independent 
generators, $\{ P_\mu, M^\mu{}_\nu, K^\mu, D \}$. 

Let us begin by making the generator $D$ for dilatation. Recall that in conformal field theory 
in the four-dimensional Minkowski space-time, the dilatation generator obeys the following algebra 
for a local primary operator $O_i (x)$ of conformal dimension $\Delta_i$ \cite{Gross, Nakayama}:
\begin{eqnarray}
[ i D, O_i (x) ] = x^\mu \partial_\mu O_i (x) + \Delta_i O_i (x).
\label{D-com}  
\end{eqnarray}
Since the scalar field $\phi (x)$ has conformal dimension $1$ as shown in (\ref{Spec-Conf-us}), it must satisfy the equation:
\begin{eqnarray}
[ i D, \phi (x) ] = x^\mu \partial_\mu \phi (x) + \phi (x).
\label{D-phi-com}  
\end{eqnarray}
 
To be consistent with this equation, we shall make a generator $D$ for the dilatation.
From the definitions (\ref{Res-gen}) and (\ref{Trans-GL}), we find
\begin{eqnarray}
[ i G^\mu{}_\mu, \phi (x) ] = x^\mu \partial_\mu \phi (x), \qquad
[ i D_0, \phi (x) ] = - \phi (x),
\label{GD-phi-com}  
\end{eqnarray}
where Eqs. (\ref{pi(B)-B}) and (\ref{dot-phi-b-eq}) were used. 
The following linear combination of $G^\mu{}_\nu$ and $D_0$ does the job:
\begin{eqnarray}
D \equiv G^\mu{}_\mu - D_0.
\label{D-def}  
\end{eqnarray}
As a consistency check, it is valuable to see how this operator $D$ acts on the metric field
whose result reads
\begin{eqnarray}
[ i D, g_{\sigma\tau} ] &=& [ i G^\mu{}_\mu, g_{\sigma\tau} ] - [ i D_0, g_{\sigma\tau} ] 
= ( x^\mu \partial_\mu g_{\sigma\tau} + 2 g_{\sigma\tau} ) - 2 g_{\sigma\tau}
\nonumber\\
&=&  x^\mu \partial_\mu g_{\sigma\tau},
\label{D-g-com}  
\end{eqnarray}
where we have used Eqs. (\ref{Ext-de-Donder}), (\ref{g-b}), (\ref{g-dot b}) and (\ref{dotB-g}).
This equation implies that the metric field is a conformal field of conformal dimension $0$ 
as expected in the conventional conformal field theory.

Next, let us calculate an algebra among the generators $\{ P_\mu, G^\mu{}_\nu, K^\mu, D \}$.
After some calculations, we find that the algebra closes and takes the form:
\begin{eqnarray}
&{}& [ P_\mu, P_\nu ] = 0, \quad 
[ P_\mu, G^\rho{}_\sigma ] = i P_\sigma \delta^\rho_\mu, \quad
[ P_\mu, K^\nu ] = - 2 i ( G^\rho{}_\rho - D ) \delta^\nu_\mu, \quad
\nonumber\\
&{}& [ P_\mu, D ] = i P_\mu, \quad 
[ G^\mu{}_\nu, G^\rho{}_\sigma ] = i ( G^\mu{}_\sigma \delta^\rho_\nu - G^\rho{}_\nu \delta^\mu_\sigma),
\nonumber\\
&{}& [ G^\mu{}_\nu, K^\rho ] = i K^\mu \delta^\rho_\nu, \quad
[ G^\mu{}_\nu, D ] = [ K^\mu, K^\nu ] = 0, 
\nonumber\\
&{}& [ K^\mu, D ] = - i K^\mu, \quad
[ D, D] = 0. 
\label{Grav-conf0}  
\end{eqnarray}
In this way, we have succeeded in showing that we have an extended conformal algebra in the formalism
at hand. As mentioned above, the essential difference between the extended conformal symmetry and
the conventional conformal symmetry is the presence of the $GL(4)$ symmetry instead of the $SO(1, 3)$
Lorentz symmetry, and this fact provides us with an important proof of exact masslessness of the graviton
\cite{NO}. 

We mention two remarks, one of which is that to show this closure we do not have to use the
gravitational ETCRs such as $[ \dot g_{\mu\nu}, g_{\rho\sigma}^\prime ]$ which is missing in the
formalism under consideration owing to an unspecified gravitational classical action. 
As the second remark, let us note that we have two different ways for deriving the algebra (\ref{Grav-conf0}).
The direct way is to use almost all the ETCRs obtained in this article. In addition to it, we need some nontrivial 
ETCRs which are not written out explicitly but can be obtained through the ETCRs appeared thus far. For instance, 
we sometimes have to make use of the ETCR:
\begin{eqnarray}
&{}& [ \dot b_\mu, \dot b_\nu^\prime ] = i \tilde f \phi^{-2} ( \delta_\nu^0 \ddot b_\mu 
+ 2 \delta_\nu^k \partial_k \dot b_\mu ) \delta^3
- 2 i \tilde f \tilde g^{0k} \partial_k [ \tilde f \phi^{-2} ( \partial_\mu b_\nu + \partial_\nu b_\mu ) \delta^3 ] 
\nonumber\\
&{}& + i \tilde f^2 \phi^{-2} ( 2 \tilde g^{0k} \delta_\nu^l - \tilde g^{kl} \delta_\nu^0 ) 
\partial_k \partial_l b_\mu \delta^3 
- i \partial_0 [ \tilde f \phi^{-2} ( \partial_\mu b_\nu + \partial_\nu b_\mu ) ] \delta^3. 
\label{dot-b-dot-b}  
\end{eqnarray}
This ETCR is obtained by taking time-derivative of the latter equation in (\ref{b-b}) and then using
the field equation (\ref{b-rho-eq}). The other easier way of the derivation of (\ref{Grav-conf0}) is 
to appeal to the $IOSp(10|10)$ algebra in Eq. (\ref{IOSp-algebra}).  To illustrate the two different ways
of derivation, we will explicitly evaluate the algebra $[ P_\mu, K^\nu ] = - 2i ( G^\rho{}_\rho - D ) \delta_\mu^\nu$
in Appendix D.

To extract conformal algebra from the extended conformal algebra (\ref{Grav-conf0}), it is necessary to introduce
the ``Lorentz generator'', which can be contructed from the $GL(4)$ generator as
\begin{eqnarray}
M_{\mu\nu} \equiv - \eta_{\mu\rho} G^\rho{}_\nu + \eta_{\nu\rho} G^\rho{}_\mu. 
\label{Lor-gene}  
\end{eqnarray}
In terms of the generator $M_{\mu\nu}$, the algebra (\ref{Grav-conf0}) can be cast to the form:
\begin{eqnarray}
&{}& [ P_\mu, P_\nu ] = 0, \quad 
[ P_\mu, M_{\rho\sigma} ] = i ( P_\rho \eta_{\mu\sigma} - P_\sigma \eta_{\mu\rho} ), 
\nonumber\\
&{}& [ P_\mu, K^\nu ] = - 2 i ( G^\rho{}_\rho - D ) \delta^\nu_\mu, \quad
[ P_\mu, D ] = i P_\mu, 
\nonumber\\
&{}& [ M_{\mu\nu}, M_{\rho\sigma} ] = - i ( M_{\mu\sigma} \eta_{\nu\rho} - M_{\nu\sigma} \eta_{\mu\rho}
+ M_{\rho\mu} \eta_{\sigma\nu} - M_{\rho\nu} \eta_{\sigma\mu}),
\nonumber\\
&{}& [ M_{\mu\nu}, K^\rho ] = i ( - K_\mu \delta^\rho_\nu + K_\nu \delta^\rho_\mu ), \quad
[ M_{\mu\nu}, D ] = [ K^\mu, K^\nu ] = 0, 
\nonumber\\
&{}& [ K^\mu, D ] = - i K^\mu, \quad
[ D, D] = 0. 
\label{Grav-conf}  
\end{eqnarray}
where we have defined $K_\mu \equiv \eta_{\mu\nu} K^\nu$. It is worthwhile to point out that
the the algebra (\ref{Grav-conf}) in quantum gravity coincides with the well-known conformal algebra 
except for the expression of $[ P_\mu, K^\nu ]$, which takes the form, $[ P_\mu, K^\nu ] 
= - 2 i ( \delta^\nu_\mu D + M_\mu{}^\nu )$ in the conformal algebra.
However, this difference reflects only the difference of the definition of conformal dimension in both gravity 
and conformal field theory, for which the metric tensor field $g_{\mu\nu}$ has $-2$ in gravity as seen 
in Eq. (\ref{Weyl-transf}) or Eq. (\ref{Spec-Conf-us}) while it has $0$ in conformal field theory as seen 
in Eq. (\ref{D-g-com}). Moreover, the absence of the term $M_\mu{}^\nu$ on the RHS of $[ P_\mu, K^\nu ]$
in (\ref{Grav-conf}) can be understood from the fact that this term comes from the quadratic terms in $x^\mu$ 
in Eq. (\ref{Spec-Conf}), which is now missing at present owing to our linear approximation as seen 
in Eq. (\ref{Spec-Conf-us}). Accordingly, in this sense the algebra (\ref{Grav-conf}) is essentially equivalent 
to the conventional conformal algebra, and thus we have succeeded in obtaining the conformal symmetry 
within the framework of quantum gravity.

\section{Spontaneous breakdown of symmetries}

In the theory under consideration, there is a huge global symmetry called the Poincar\'e-like $IOSp(10|10)$
symmetry where we have $20$ $P^M$ generators and $200$ $M^{MN}$ generators so totally $220$ symmetry generators. 
Then, it is valuable to investigate which symmetries are spontaneously broken or survive even in quantum regime. 
In particular, since our world has a built-in scale in it, a global scale symmetry or dilatation must be spontaneously broken
if our theory makes sense.

In order to clarify the mechanism of symmetry breakdown, let us first postulate the existence of a unique vacuum 
$| 0 \rangle$, which is normalized to be the unity:
\begin{eqnarray}
\langle 0 | 0 \rangle = 1.
\label{Vac-norm}  
\end{eqnarray}
 It is also natural to assume that the vacuum is translation invariant
\begin{eqnarray}
P_\mu | 0 \rangle \equiv P_\mu (b) | 0 \rangle = 0.
\label{Trans-Vac}  
\end{eqnarray}
Furthermore, it is supposed that the vacuum expectation values (VEVs) of the metric tensor $g_{\mu\nu}$ and 
the scalar field $\phi$ are respectively the Minkowski metric $\eta_{\mu\nu}$ and a non-zero constant $\phi_0 \neq 0$:
\begin{eqnarray}
\langle 0 | g_{\mu\nu} | 0 \rangle = \eta_{\mu\nu}, \qquad
\langle 0 | \phi | 0 \rangle = \phi_0. 
\label{VEV-A}  
\end{eqnarray}

With these assumptions, a straightforward calculation reveals the following nonvanishing VEVs for $P^M$
\begin{eqnarray}
&{}& \langle 0 | [ i P^\mu (x), b_\rho ] | 0 \rangle = - \delta^\mu_\rho, 
\nonumber\\
&{}& \langle 0 | [ i P (\sigma), B ] | 0 \rangle = - \langle 0 | [ i P (B), \sigma ] | 0 \rangle = 1,
\nonumber\\
&{}& \langle 0 | \{ i P^\mu (c^\tau), \bar c_\rho \} | 0 \rangle = i \delta^\mu_\rho, \quad
\langle 0 | \{ i P_\mu (\bar c_\tau), c^\rho \} | 0 \rangle = - i \delta^\rho_\mu, 
\nonumber\\
&{}& \langle 0 | \{ i P (c), \bar c \} | 0 \rangle = - \langle 0 | \{ i P (\bar c), c \} | 0 \rangle = i, 
\label{P-VEV}  
\end{eqnarray}
and the ones for $M^{MN}$
\begin{eqnarray}
&{}& \langle 0 | [ i M^{\mu\nu} (x, x), \frac{1}{2} ( \partial_\lambda b_\rho 
- \partial_\rho b_\lambda ) ] | 0 \rangle = - ( \delta^\mu_\lambda \delta^\nu_\rho
- \delta^\nu_\lambda \delta^\mu_\rho ),
\nonumber\\
&{}& \langle 0 | [ i M^\mu{}_\nu (x, b), g_{\sigma\tau} ] | 0 \rangle = \delta^\mu_\sigma \eta_{\nu\tau}
+ \delta^\mu_\tau \eta_{\nu\sigma},
\nonumber\\
&{}& \langle 0 | [ i M^\mu (x, \sigma), \partial_\lambda B ] | 0 \rangle 
= \langle 0 | [ i M^\mu (x, B), \partial_\lambda \sigma ] | 0 \rangle = \delta^\mu_\lambda,
\nonumber\\
&{}& \langle 0 | \{ i M^{\mu\nu} (x, c^\tau), \partial_\lambda \bar c_\rho \} | 0 \rangle 
= i \delta^\mu_\lambda \delta^\nu_\rho,   \quad
\langle 0 | \{ i M^\mu{}_\nu (x, \bar c_\tau), \partial_\lambda c^\rho \} | 0 \rangle 
= - i \delta^\mu_\lambda \delta^\rho_\nu, 
\nonumber\\
&{}& \langle 0 | \{ i M^\mu (x, c), \partial_\lambda \bar c \} | 0 \rangle = i \delta^\mu_\lambda,   \quad
\langle 0 | \{ i M^\mu (x, \bar c), \partial_\lambda c \} | 0 \rangle = - i \delta^\mu_\lambda, 
\nonumber\\
&{}& \langle 0 | [ i M (\sigma, B), \sigma ] | 0 \rangle = \sigma_0, 
\nonumber\\
&{}& \langle 0 | \{ i M^\mu (\sigma, c^\tau), \bar c_\rho \} | 0 \rangle = i \sigma_0 \delta^\mu_\rho,   \quad
\langle 0 | \{ i M_\mu (\sigma, \bar c_\tau), c^\rho \} | 0 \rangle = - i \sigma_0 \delta^\rho_\mu, 
\nonumber\\
&{}& \langle 0 | \{ i M (\sigma, c), \bar c \} | 0 \rangle = - \langle 0 | \{ i M (\sigma, \bar c), c \} | 0 \rangle 
= i \sigma_0, 
\label{M-VEV}  
\end{eqnarray}
where $\langle 0 | \sigma(x) | 0 \rangle \equiv \sigma_0$. These VEVs clearly show that the symmetries generated by the conserved charges 
\begin{eqnarray*}
&{}& \{ P^\mu (x), P (\sigma), P (B), P^\mu (c^\tau), P_\mu (\bar c_\tau), P (c), P (\bar c), M^{\mu\nu} (x, x), 
M^\mu{}_\nu (x, b),
\nonumber\\
&{}& M^\mu (x, \sigma), M^\mu (x, B), M^{\mu\nu} (x, c^\tau), M^\mu{}_\nu (x, \bar c_\tau), M^\mu (x, c), M^\mu (x, \bar c),
\nonumber\\
&{}& M (\sigma, B), M^\mu (\sigma, c^\tau), M_\mu (\sigma, \bar c_\tau), M (\sigma, c), M (\sigma, \bar c) \}
\label{SSB-charge}  
\end{eqnarray*}
are necessarily broken spontaneously, thereby $g_{\mu\nu}, b_\mu, \sigma, B, c^\mu, \bar c_\mu, c$ and $\bar c$ acquiring massless 
Nambu-Goldstone modes. 
On the other hand, the remaining generators
\begin{eqnarray*}
&{}& \{ M_{\mu\nu} (b, b), M_\mu (b, \sigma), M_\mu (b, B), M_\mu{}^\nu (b, c^\tau), M_{\mu\nu} (b, \bar c_\tau),
M_\mu (b, c), M_\mu (b, \bar c),  
\nonumber\\
&{}& M^\mu (B, c^\tau), M_\mu (B, \bar c_\tau), M (B, c), M(B, \bar c), M^{\mu\nu} (c^\tau, c^\lambda), 
M^\mu{}_\nu (c^\tau, \bar c_\lambda), 
\nonumber\\
&{}& M^\mu (c^\tau, c), M^\mu (c^\tau, \bar c), M_{\mu\nu} (\bar c_\tau, \bar c_\lambda), M_\mu (\bar c_\tau, c), 
M_\mu (\bar c_\tau, \bar c), 
\nonumber\\
&{}& M (c, c), M (c, \bar c), M (\bar c, \bar c) \},
\label{No-SSB-charge}  
\end{eqnarray*}
are found to be unbroken.

We are now ready to show that $GL(4)$, special conformal symmetry and dilatation are spontaneously broken down to 
the Poincar\'e symmetry. For this aim, it is more convenient to utilize the extended conformal algebra (\ref{Grav-conf0}) 
rather than the Poincar\'e-like $IOSp(10|10)$ algebra (\ref{IOSp-algebra}) and the conformal algebra (\ref{Grav-conf}).
It is then easy to see that the VEV of a commutator between the $GL(4)$ generator and the metric field reads     
\begin{eqnarray}
\langle 0 | [ i G^\mu{}_\nu, g_{\sigma\tau} ] | 0 \rangle 
= \delta^\mu_\sigma \eta_{\nu\tau} + \delta^\mu_\tau \eta_{\nu\sigma}.
\label{G-g-CM}  
\end{eqnarray}
Thus, the Lorentz generator, which is defined in Eq. (\ref{Lor-gene}), has the vanishing VEV:
\begin{eqnarray}
\langle 0 | [ i M_{\mu\nu}, g_{\sigma\tau} ] | 0 \rangle = 0.
\label{M-g-CM}  
\end{eqnarray}
On the other hand, the symmetric part, which is defined as $\bar M_{\mu\nu} \equiv \eta_{\mu\rho}
G^\rho{}_\nu + \eta_{\nu\rho} G^\rho{}_\mu$, has the non-vanishing VEV: 
\begin{eqnarray}
\langle 0 | [ i \bar M_{\mu\nu}, g_{\sigma\tau} ] | 0 \rangle = 2 ( \eta_{\mu\sigma} \eta_{\nu\tau}
+ \eta_{\mu\tau} \eta_{\nu\sigma} ).
\label{BM-g-CM}  
\end{eqnarray}
Thus, the $GL(4)$ symmetry is spontaneously broken to the Lorentz symmetry where the corresponding 
Nambu-Goldstone boson with ten independent components is nothing but the massless graviton \cite{NO}.
Here it is interesting that in a sector of the scalar field, the $GL(4)$ symmetry and the Lorentz
symmetry as well, do not give rise to a symmetry breaking as can be checked in the commutators:
\begin{eqnarray}
\langle 0 | [ i G^\mu{}_\nu, \phi ] | 0 \rangle  = \langle 0 | [ i M_{\mu\nu}, \phi ] | 0 \rangle 
= \langle 0 | [ i \bar M_{\mu\nu}, \phi ] | 0 \rangle = 0.
\label{G-phi-CM}  
\end{eqnarray}
 
Next, we wish to clarify how the dilatation and special conformal symmetry are spontaneously broken and
what the corresponding Nambu-Goldstone bosons are. As for the dilatation, it is not the gravitational
field but the dilaton that triggers a spontaneous symmetry breaking since for the metric 
Eq. (\ref{D-g-com}) provides us with 
\begin{eqnarray}
\langle 0 | [ i D, g_{\sigma\tau} ] | 0 \rangle = 0,
\label{VEV-D-g}  
\end{eqnarray}
while, for the dilaton, from Eq. (\ref{D-phi-com}) we have
\begin{eqnarray}
\langle 0 | [ i D, \sigma ] | 0 \rangle = 1.
\label{VEV-D-sigma}  
\end{eqnarray}
This fact elucidates the spontaneous symmetry breakdown of the dilatation whose
Nambu-Goldstone boson is just the massless dilaton $\sigma(x)$. 

Regarding the special conformal symmetry, we find
\begin{eqnarray}
\langle 0 | [ i K^\mu, \partial_\nu \sigma ] | 0 \rangle = 2 \delta^\mu_\nu.
\label{VEV-K-phi}  
\end{eqnarray}
This equation means that the special conformal symmetry is certainly broken spontaneously
and its Nambu-Goldstone boson is the derivative of the dilaton. This interpretation can be
also verified from the extended conformal algebra. In the algebra (\ref{Grav-conf0}), we have
a commutator between $P_\mu$ and $K^\nu$:
\begin{eqnarray}
[ P_\mu, K^\nu ] = - 2 i ( G^\rho{}_\rho - D ) \delta^\nu_\mu.
\label{P-K}  
\end{eqnarray}
Let us consider the Jacobi identity:
\begin{eqnarray}
[ [ P_\mu, K^\nu ], \sigma ] + [ [ K^\nu, \sigma ], P_\mu ] + [ [ \sigma, P_\mu ], K^\nu ] = 0.
\label{Jacobi}  
\end{eqnarray}
Using the translational invariance of the vacuum in Eq. (\ref{Trans-Vac}) and the equation
\begin{eqnarray}
[ P_\mu, \sigma ] =  - i \partial_\mu \sigma,
\label{Jacobi2}  
\end{eqnarray}
and taking the VEV of the Jacobi identity (\ref{Jacobi}), we can obtain the VEV
\begin{eqnarray}
\langle 0 | [ K^\nu, \partial_\mu \sigma ] | 0 \rangle &=& - 2 \delta^\nu_\mu
\langle 0 | [ G^\rho{}_\rho - D, \sigma ] | 0 \rangle
\nonumber\\
&=& - 2 i \delta^\nu_\mu,
\label{VEV-Jacobi}  
\end{eqnarray}
which coincides with Eq. (\ref{VEV-K-phi}) as promised. In other words, the $GL(4)$ 
symmetry is spontaneously broken to the Poincar\'e symmetry whose Nambu-Goldstone
boson is the graviton, the dilatation and the special conformal symmetry are at the
same time spontaneously broken and the corresponding Nambu-Goldstone bosons are the dilaton
and the derivative of the dilaton, respectively. It is not surprising that this pattern of the symmetry breaking 
for the dilatation and special conformal symmetry is the same as that of the conventional conformal field 
theory \cite{Kobayashi} since we share the same conformal algebra.

\section{Conclusion}

In this article, we have performed a manifestly covariant quantization and contructed a quantum 
theory of a general gravitational theory, which is invariant under both general coordinate transformation
(GCT) and Weyl transformation, within the framework of the BRST formalism. In the present formulation,
since we have started with such a general action, it is impossible to derive a concrete expression of the Einstein's equation 
as well as equal-time commutation relations (ETCRs) relevant to the metric tensor and its derivatives such as 
$[ \dot g_{\mu\nu}, g_{\rho\sigma}^\prime ]$. 
The key point is, however, that without the Einstein's equation and those ETCRs one can derive the field equation 
for $b_\mu$ and the other nontrivial ETCRs, in particular, $[ \dot b_\mu, b_\nu^\prime ]$, which have been so far derived 
from the Einstein's equation and the metric's ETCRs \cite{N-O-text}, by the help of two kinds of BRST transformations 
and field equations except for the Einstein's equation.    

Although we do not fix the classical gravitational action and consider a general one, it is remarkable that we have
a huge global symmetry, called a Poincar\'e-like $IOSp(10|10)$ global transformation with $220$ symmetry generators, 
which is larger than an $ISOp(8|8)$ with $144$ symmetry generators in case of general relativity \cite{N-O-text}, owing to 
the presence of the Weyl symmetry in our formulation. In a flat Minkowski background, the Poincar\'e-like $IOSp(10|10)$ 
symmetry naturally includes the extended conformal algebra where the usual Lorentz symmetry is extended to a
larger $GL(4)$ symmetry. Then, it is obvious that this extended conformal algebra includes conformal algebra as a
subgroup. In other words, in our formulation of quantum gravity, the conformal symmetry emerges from not the classical action 
but the gauge-fixing and FP ghost actions. 

According to the Zumino theorem \cite{Zumino}, the theories which are invariant under the GCT and Weyl transformation,
have conformal invariance in the flat Minkowski background at the classical level.  The present study partially supports 
a conjecture that the Zumino theorem could be valid even in quantum gravity. However, there is a loophole in our argument.   
In this article, we have assumed that a classical action does not involve more than first order derivatives of the metric and 
matter fields. If the classical theory includes the higher-derivative terms as in quadratic gravity, we have to rewrite the theory 
in order not to include more than first order derivatives by introducing suitabe auxiliary fields when we perform the canonical 
quantization.  Then, the introduction of the auxiliary fields sometimes gives rise to a new local symmetry such as the St\"{u}ckelberg   
symmetry \cite{Kimura1, Kimura2, Kimura3, Kubo1, Kubo2, Oda-Ohta}. In such a situation, our argument is not valid. Thus, we can
address a generalized Zumino theorem in quantum gravity as follows: The Zumino theorem is certainly valid in quantum gravity
when a classical action does not involve more than first order derivatives.
  
As future works, we would like to list up two important issues. First, it is well-known that Weyl symmetry is explicitly broken 
by conformal anomaly at the quantum level \cite{Duff}. However, since a scalar field is now at our disposal, it is not obvious that
the Weyl symmetry is anomalous or not in our theory. Actually, in a Weyl invariant scalar-tensor gravity, there is no conformal anomaly
\cite{Englert}.  

Another important issue is related to the non-renormalizability of general relativity. To remedy the problem,
we are accustomed to adding quadratic terms in the Riemann curvatures and consider the Einstein-Hilbert term supplemented 
with the terms containing $R^2$ and $R_{\mu\nu} R^{\mu\nu}$, but then we meet another serious problem of non-unitarity. 
We conjecture that the non-unitarity problem in the higher-derivative gravities might be solved by a mechanism of ghost confinement
\cite{Kimura-conf} and then the Poincar\'e-like global symmetry would play a critical role, as in the confinement of quarks and gluons
by the BRST symmetry in QCD \cite{Kugo-conf}. We shall return these two issues in near future.

\section*{Acknowledgment}

This work is supported in part by the JSPS Kakenhi Grant No. 21K03539.

\appendix
\addcontentsline{toc}{section}{Appendix~\ref{app:scripts}: Training Scripts}
\section*{Appendix}
\label{app:scripts}
\renewcommand{\theequation}{A.\arabic{equation}}
\setcounter{equation}{0}

\section{$GL(4)$ invariance of the extended de Donder gauge condition}
\def\T{\text{T}}

In this appendix, we present a proof of the global $GL(4)$ invariance of the extended de Donder gauge condition 
(\ref{Ext-de-Donder}). To do that, let us first recall the $GL(4)$ transformation:
\begin{eqnarray}
x^{\prime\mu} = A^\mu{}_\nu x^\nu,
\label{GL4-x}  
\end{eqnarray}
where $A^\mu{}_\nu$ is a constant and invertible $4 \times 4$ matrix belonging to a $GL(4, R)$.
From this definition, $A^\mu{}_\nu$ and its inverse $A^{-1 \mu}{}_\nu$ can be expressed as
\begin{eqnarray}
A^\mu{}_\nu = \frac{\partial x^{\prime\mu}}{\partial x^\nu},  \qquad
A^{-1 \mu}{}_\nu = \frac{\partial x^\mu}{\partial x^{\prime\nu}}.
\label{A-matrix}  
\end{eqnarray}
Under the $GL(4)$ transformation, the metric and the scalar field transform as
\begin{eqnarray}
&{}& g^\prime_{\mu\nu} (x^\prime) = A^{-1 \rho}{}_\mu A^{-1 \sigma}{}_\nu g_{\rho\sigma} (x), \qquad
g^{\prime\mu\nu} (x^\prime) = A^\mu{}_\rho A^\nu{}_\sigma g^{\rho\sigma} (x),
\nonumber\\
&{}& \sqrt{- g^\prime (x^\prime)} = (\det A)^{-1} \sqrt{- g (x)}, \qquad
\phi^\prime (x^\prime) = \phi (x).
\label{GL4-g}  
\end{eqnarray}

Using the transformation law and $\partial^\prime_\mu = A^{-1 \nu}{}_\mu \partial_\nu$, we can show that
\begin{eqnarray}
\partial^\prime_\mu ( \sqrt{- g^\prime (x^\prime) } g^{\prime\mu\nu} (x^\prime) \phi^{\prime 2} (x^\prime) )
= (\det A)^{-1} A^\nu{}_\sigma \partial_\rho ( \sqrt{- g (x)} g^{\rho\sigma} (x) \phi^2 (x) ).
\label{GL4-gauge}  
\end{eqnarray}
Thus, the extended de Donder gauge condition (\ref{Ext-de-Donder}) is invariant under the $GL(4)$
transformation in the sense that when the extended de Donder gauge condition is imposed in a frame
it holds true in a $GL(4)$ transformed frame as well. In a similar manner, the $GL(4)$ invariance of 
the unitary gauge $\phi = \rm{constant}$ and the Lorenz gauge $\nabla_\mu S^\mu = 0$ can be
shown. Precisely speaking, the extended de Donder gauge condition is transformed as a vector density 
under the $GL(4)$ transformation as seen in Eq. (\ref{GL4-gauge}).

Indeed, it is a gauge-fixing action that is exactly invariant under the $GL(4)$ transformation. To see this fact,
let us consider a gauge-fixed action corresponding to the extended de Donder gauge condition (\ref{Ext-de-Donder}):
\begin{eqnarray}
S_{GF} = - \int d^4 x \sqrt{-g} g^{\mu\nu} \phi^2 \partial_\mu b_\nu.
\label{GF-action}  
\end{eqnarray}
Under the $GL(4)$ transformation, this action transforms as
\begin{eqnarray}
S_{GF}^\prime &=& - \int d^4 x^\prime \sqrt{-g^\prime} g^{\prime\mu\nu} \phi^{\prime 2} \partial_\mu^\prime b_\nu^\prime
\nonumber\\
&=&  - \int d^4 x \det A (\det A)^{-1} A^\mu{}_\rho A^\nu{}_\sigma \sqrt{-g} g^{\rho\sigma} \phi^2 
A^{-1 \tau}{}_\mu \partial_\tau ( A^{-1 \lambda}{}_\nu b_\lambda )
\nonumber\\
&=& S_{GF},
\label{GF-action2}  
\end{eqnarray}
where we have used
\begin{eqnarray}
b_\mu^\prime (x^\prime) = A^{-1 \nu}{}_\mu b_\nu (x),  \qquad
d^4 x^\prime = \det A \, d^4 x. 
\label{d4-x}  
\end{eqnarray}
Hence, the gauge-fixed action is precisely invariant under the $GL(4)$ transformation.

\renewcommand{\theequation}{B.\arabic{equation}}
\setcounter{equation}{0}

\section{Proof of Eqs. (\ref{dot g-b}) and (\ref{g-dot b})}

In this Appendix, we derive Eqs. (\ref{dot g-b}) and (\ref{g-dot b}) by following the method developed previously in case of 
the de Donder gauge condition \cite{Oda-Q}.

The translational invariance in general requires the validity of the following equation for a generic field $\Phi(x)$:
\begin{eqnarray}
[ \Phi(x), P_\rho ] = i \partial_\rho \Phi(x),
\label{Translation-eq}  
\end{eqnarray}
where $P_\rho$ is the generator of the translation which is now given by
\begin{eqnarray}
P_\rho = \int d^3 x \, \tilde g^{0\lambda} \phi^2 \partial_\lambda b_\rho.
\label{Def-P}  
\end{eqnarray}

Now let us consider the specific case $\Phi(x) = g_{\mu\nu}(x)$
\begin{eqnarray}
[ g_{\mu\nu}(x), P_\rho ] = [ g_{\mu\nu}(x), \int d^3 x' \, \tilde g^{0\lambda\prime} \phi^{2\prime} 
\partial_\lambda b_\rho^\prime ] = i \partial_\rho g_{\mu\nu}(x).
\label{g-translation}  
\end{eqnarray}  
Taking $x^0 = x^{\prime0}$ and using $[ g_{\mu\nu}, \tilde g^{0\lambda \prime} \phi^{2\prime} ] = 0$, we have
\begin{eqnarray}
\int d^3 x' \, \tilde g^{0\lambda}(x^\prime) \phi^2 (x^\prime) [ g_{\mu\nu}, \partial_\lambda b_\rho^\prime ] 
= i \partial_\rho g_{\mu\nu}(x).
\label{g-translation2}  
\end{eqnarray}  
Using the extended de Donder gauge condition (\ref{Ext-de-Donder}) and Eq. (\ref{g-b}), this equation can be rewritten as
\begin{eqnarray}
\int d^3 x' \, \tilde g^{00} (x^\prime) \phi^2 (x^\prime) [ g_{\mu\nu}, \dot b_\rho^\prime ] 
= i \left[ \partial_\rho g_{\mu\nu} + \partial_0 ( \tilde g^{00} \phi^2 ) \tilde f \phi^{-2} ( \delta_\mu^0 g_{\rho\nu}
+ \delta_\nu^0 g_{\rho\mu} ) \right],
\label{g-translation3}  
\end{eqnarray}  
which is easily solved for $[ g_{\mu\nu}, \dot b_\rho^\prime ]$ to be
\begin{eqnarray}
[ g_{\mu\nu}, \dot b_\rho^\prime ] 
&=& i \tilde f  \phi^{-2} \left[ \partial_\rho g_{\mu\nu} + \partial_0 ( \tilde g^{00} \phi^2 ) \tilde f \phi^{-2} ( \delta_\mu^0 g_{\rho\nu}
+ \delta_\nu^0 g_{\rho\mu} ) \right] \delta^3 
\nonumber\\
&+& F_{(\mu\nu)\rho}{}^k \partial_k (\tilde f \phi^{-2} \delta^3),
\label{g-dot-b-sol}  
\end{eqnarray}  
where $F_{(\mu\nu)\rho}{}^k$ is an arbitrary function which is symmetric under the exchange of $\mu
\leftrightarrow \nu$.

Next, to fix the function $F_{(\mu\nu)\rho}{}^k$, let us take account of the consistency with the extended de Donder 
gauge condition (\ref{Ext-de-Donder}):
\begin{eqnarray}
[ \partial_\mu ( \tilde g^{\mu\nu} \phi^2 ), b_\rho^\prime ] = 0.
\label{Consist-Donder}  
\end{eqnarray}  
An explicit calculation reveals that Eq. (\ref{Consist-Donder}) leads to an equation for $F_{(\mu\nu)\rho}{}^k$:
\begin{eqnarray}
\left( \tilde g^{0\alpha} g^{\nu\beta} - \frac{1}{2} \tilde g^{0\nu} g^{\alpha\beta} \right) F_{(\alpha\beta)\rho}{}^k
= - i ( \tilde g^{0k} \delta_\rho^\nu + \tilde g^{0\nu} \delta_\rho^k - \tilde g^{k\nu} \delta_\rho^0 ).
\label{Consist-Donder2}  
\end{eqnarray}  
This equation has the unique solution given by 
\begin{eqnarray}
F_{(\mu\nu)\rho}{}^k
= i [ ( \delta_\mu^k - 2 \delta_\mu^0 \tilde f \tilde g^{0k} ) g_{\rho\nu} + (\mu \leftrightarrow \nu) ].
\label{F-sol}  
\end{eqnarray}  
We can therefore obtain
\begin{eqnarray}
[ g_{\mu\nu}, \dot b_\rho^\prime ] 
&=& i \Bigl\{ [ \tilde f \phi^{-2} \partial_\rho g_{\mu\nu} - \partial_0 ( \tilde f \phi^{-2} ) ( \delta_\mu^0 g_{\rho\nu}
+ \delta_\nu^0 g_{\rho\mu} ) ] \delta^3 
\nonumber\\
&+& [ ( \delta_\mu^k - 2 \delta_\mu^0 \tilde f \tilde g^{0k} ) g_{\rho\nu} + (\mu \leftrightarrow \nu) ]
\partial_k (\tilde f \phi^{-2} \delta^3) \Bigr\},
\label{g-dot-b-final}  
\end{eqnarray}  
which is nothing but the former equation in Eq. (\ref{g-dot b}). The latter equation in Eq. (\ref{g-dot b})
can be easily obtained from Eq. (\ref{g-dot-b-final}).
Finally, using Eqs. (\ref{identity}) and (\ref{g-b}), we can arrive at another 
equation (\ref{dot g-b}). It is of interest that Eq. (\ref{dot g-b}) can be derived from only the translational invariance and 
the extended de Donder gauge condition without reference to the classical Lagrangian ${\cal{L}}_c$ which knows 
information of the dynamics of the gravitational field $g_{\mu\nu}$ and the scalar field $\phi$.

\renewcommand{\theequation}{C.\arabic{equation}}
\setcounter{equation}{0}

\section{Proof of $[ \dot b_\mu, b_\nu^\prime ] = - i \tilde f \phi^{-2} ( \partial_\mu b_\nu + \partial_\nu b_\mu ) \delta^3$}

In this Appendix, we present a proof of the ETCR, $[ \dot b_\mu, b_\nu^\prime ] = - i \tilde f \phi^{-2} ( \partial_\mu b_\nu 
+ \partial_\nu b_\mu ) \delta^3$ on the basis of the BRST transformation without using the Einstein's equation.
This calculation exhibits that the BRST transformation offers a very powerful method for deriving a nontrivial ETCR.  

Let us begin by the ETCR in Eq. (\ref{dot-c-b-eq}):
\begin{eqnarray}
[ \dot{\bar c}_\mu, b_\nu^\prime ] = - i \tilde f \phi^{-2}  \partial_\nu \bar c_\mu \delta^3.
\label{Start}  
\end{eqnarray} 
Then, taking its GCT BRST transformation yields 
\begin{eqnarray}
[ i \dot B_\mu, b_\nu^\prime ] - \{ \dot{\bar c}_\mu, - c^{\alpha\prime} \partial_\alpha b_\nu^\prime \}
&=& - i ( \delta_B \tilde f \phi^{-2} - 2 \tilde f \phi^{-3} \delta_B \phi ) \partial_\nu \bar c_\mu \delta^3
\nonumber\\
&-& i \tilde f \phi^{-2} \partial_\nu ( i B_\mu ) \delta^3.
\label{Next}  
\end{eqnarray} 

Using Eq. (\ref{b-rho-field}), the first term on the left-hand side (LHS) and the last term
on the right-hand side (RHS) can be rewritten as
\begin{eqnarray}
[ i \dot B_\mu, b_\nu^\prime ] &=& i [ \dot b_\mu, b_\nu^\prime ] 
- [ \partial_0 ( c^\alpha \partial_\alpha \bar c_\mu ), b_\nu^\prime ],
\nonumber\\
- i \tilde f \phi^{-2} \partial_\nu ( i B_\mu ) \delta^3 &=& f \phi^{-2} \partial_\nu b_\mu \delta^3
+ i \tilde f \phi^{-2} \partial_\nu ( c^\alpha \partial_\alpha \bar c_\mu ) \delta^3.
\label{1st-last}  
\end{eqnarray} 
Furthermore, we can calculate $\delta_B \tilde f$ to be
\begin{eqnarray}
\delta_B \tilde f = - \tilde f^2 \delta_B \tilde g^{00} = \tilde f \partial_\rho c^\rho - \partial_\rho \tilde f c^\rho
- 2 \tilde f^2 \tilde g^{0\rho} \partial_\rho c^0.
\label{delta-f}  
\end{eqnarray} 
Using these equations and $\delta_B \phi = - c^\rho \partial_\rho \phi$, Eq. (\ref{Next}) can be further 
cast to the form:
\begin{eqnarray}
&{}& [ \dot b_\mu, b_\nu^\prime ] = - i [ \partial_0 ( c^\alpha \partial_\alpha \bar c_\mu ), b_\nu^\prime ]
+ i \{ \dot{\bar c}_\mu, c^{\alpha\prime} \partial_\alpha b_\nu^\prime \}
- i f \phi^{-2} \partial_\nu b_\mu \delta^3 
\nonumber\\
&{}& + \tilde f \phi^{-2} \partial_\nu ( c^\alpha \partial_\alpha \bar c_\mu ) \delta^3
- \tilde f \phi^{-2} ( \partial_\rho c^\rho - \tilde f^{-1} \partial_\rho \tilde f c^\rho
- 2 \tilde f \tilde g^{0\rho} \partial_\rho c^0 
\nonumber\\
&{}& + 2 \phi^{-1} c^\rho \partial_\rho \phi )
\partial_\nu \bar c_\mu \delta^3.
\label{Next2}  
\end{eqnarray} 
  
The remaining work is to calculate two (anti)commutators on the RHS.  It is straightforward
to carry out such calculations by using various ETCRs obtained thus far. In particular,
we make use of the following ETCRs
\begin{eqnarray}
&{}& [ \ddot{\bar c}_\mu, b_\nu^\prime ] = - i \tilde f \phi^{-2} \delta_\nu^0 \ddot{\bar c}_\mu \delta^3 
- 2 i \tilde f [ \phi^{-2} \delta_\nu^k \partial_k \dot{\bar c}_\mu \delta^3
- \tilde g^{0k} \partial_k ( \tilde f \phi^{-2} \partial_\nu \bar c_\mu \delta^3 ) ]
\nonumber\\
&{}& - i \tilde f^2 \phi^{-2} ( 2 \tilde g^{0k} \delta_\nu^l - \tilde g^{kl} \delta_\nu^0 ) 
\partial_k \partial_l \bar c_\mu \delta^3,
\label{ddot-barc-b}  
\end{eqnarray} 
and 
\begin{eqnarray}
[ \dot{\bar c}_\mu, \dot b_\nu^\prime ] = - [ \ddot{\bar c}_\mu, b_\nu^\prime ] 
- i \partial_0 ( \tilde f \phi^{-2} \partial_\nu \bar c_\mu ) \delta^3.
\label{dot-barc-dot-b}  
\end{eqnarray} 
Eq. (\ref{ddot-barc-b}) can be obtained from Eqs. (\ref{3-g-b}), (\ref{Phi-b}), (\ref{ddot-bar-c}) and
(\ref{dot-c-b-eq}). 

As a result, the two (anti)commutators on the RHS in Eq. (\ref{Next2}) are given by
\begin{eqnarray}
&{}& - i [ \partial_0 ( c^\alpha \partial_\alpha \bar c_\mu ), b_\nu^\prime ] 
= - \tilde f \phi^{-2} ( \partial_\nu c^\alpha \partial_\alpha \bar c_\mu
+ \dot c^0 \partial_\nu \bar c_\mu + \delta_\nu^0 c^0 \ddot{\bar c}_\mu ) \delta^3
\nonumber\\
&{}& - 2 \tilde f c^0 [ \phi^{-2} \delta_\nu^k \partial_k \dot{\bar c}_\mu \delta^3
- \tilde g^{0k} \partial_k ( \tilde f \phi^{-2} \partial_\nu \bar c_\mu \delta^3 ) ]
\nonumber\\
&{}& - \tilde f^2 \phi^{-2} c^0 ( 2 \tilde g^{0k} \delta_\nu^l - \tilde g^{kl} \delta_\nu^0 ) 
\partial_k \partial_l \bar c_\mu \delta^3
- c^k \partial_k ( \tilde f \phi^{-2} \partial_\nu \bar c_\mu \delta^3 ),
\label{First-CT}  
\end{eqnarray} 
and 
\begin{eqnarray}
&{}& i \{ \dot{\bar c}_\mu, c^{\alpha\prime} \partial_\alpha b_\nu^\prime \} 
= - i \tilde f \phi^{-2} \partial_\mu b_\nu \delta^3 + \tilde f \phi^{-2} \delta_\nu^0 c^0 \ddot{\bar c}_\mu \delta^3 
\nonumber\\
&{}& + 2 \tilde f [ \phi^{-2} \delta_\nu^k c^0 \partial_k \dot{\bar c}_\mu \delta^3
- \tilde g^{0k} \partial_k ( \tilde f \phi^{-2} c^0 \partial_\nu \bar c_\mu \delta^3 ) ]
\nonumber\\
&{}& + \tilde f^2 \phi^{-2} ( 2 \tilde g^{0k} \delta_\nu^l - \tilde g^{kl} \delta_\nu^0 ) 
c^0 \partial_k \partial_l \bar c_\mu \delta^3
\nonumber\\
&{}& - c^0 \partial_0 ( \tilde f \phi^{-2} \partial_\nu \bar c_\mu ) \delta^3 
+ \tilde f \phi^{-2} \partial_k ( c^k \delta^3 ) \partial_\nu \bar c_\mu.
\label{Second-CT}  
\end{eqnarray} 
Substituting Eqs. (\ref{First-CT}) and (\ref{Second-CT}) into Eq. (\ref{Next2}), we can reach 
the desired equation:
\begin{eqnarray}
[ \dot b_\mu, b_\nu^\prime ] = - i \tilde f \phi^{-2} ( \partial_\mu b_\nu + \partial_\nu b_\mu ) \delta^3.
\label{Final}  
\end{eqnarray} 

\renewcommand{\theequation}{D.\arabic{equation}}
\setcounter{equation}{0}

\section{Two derivations of $[ P_\mu, K^\nu ] = - 2i ( G^\rho{}_\rho - D ) \delta_\mu^\nu$}

For illustrative purposes, in this appendix we present two different derivations for one of conformal algebra, 
$[ P_\mu, K^\nu ] = - 2i ( G^\rho{}_\rho - D ) \delta_\mu^\nu$, one of which relies on the $IOSp(10|10)$ algebra 
in Eq. (\ref{IOSp-algebra}). Another derivation is a direct calculation based on the ETCRs obtained in this article.
We find that the former derivation method is much easier than that of the latter and this fact elucidates the power
of the Poincar\'e-like $IOSp(10|10)$ algebra.

For the translation generator $P_\mu$ in (\ref{Trans-GL}) and the special conformal generator $K^\mu$ 
in (\ref{Res-gen}), the $IOSp(10|10)$ algebra Eq. (\ref{IOSp-algebra}) provides us with
\begin{eqnarray}
[ P_\mu, K^\nu ] &=& - [ K^\nu, P_\mu ] = - 2 [ M^\nu (x, B), P_\mu (b) ] 
= - 2 [ M^{x^\nu B}, P_{b_\mu} ]
\nonumber\\
&=& - 2 i [ P^{x^\nu} \tilde \eta^B{}_{b_\mu} - (-)^{|B| |b_\mu|} P^B \tilde \eta^{x^\nu}{}_{b_\mu} ]
= 2 i P^B \delta_\mu^\nu 
\nonumber\\
&=& 2 i P (B) \delta_\mu^\nu = - 2 i D_0 \delta_\mu^\nu
= - 2i ( G^\rho{}_\rho - D ) \delta_\mu^\nu,
\label{Poin-Start}  
\end{eqnarray} 
where we have used $\tilde \eta^B{}_{b_\mu} = 0$ and $\tilde \eta^{x^\nu}{}_{b_\mu} = \delta_\mu^\nu$
at the fifth equality. Note that this derivation is very transparent.

As an alternative derivation of this algebra, one can rely on the ETCRs obtained thus far.
This derivation method is very direct but it is a bit complicated since we have to use
the extended de Donder gauge condition (\ref{Ext-de-Donder}), field equations and the integration 
by parts etc. repeatedly.  

From Eqs. (\ref{Res-gen}) and (\ref{Trans-GL}), we have
\begin{eqnarray}
&{}&  [ P_\mu, K^\nu ] = \int d^3 x d^3 x^\prime [ \tilde g^{0\lambda} \phi^2 \partial_\lambda b_\mu,
2 \tilde g^{0\rho\prime} \phi^{2\prime} x^{\nu\prime} \overset{\leftrightarrow}{\partial}_\rho B^\prime ]
\nonumber\\
&{}& = 2 \int d^3 x d^3 x^\prime \Bigl( x^{\nu\prime} [ \tilde g^{0\lambda} \phi^2 \partial_\lambda b_\mu,
\tilde g^{0\rho\prime} \phi^{2\prime} \partial_\rho B^\prime ]
- [ \tilde g^{0\lambda} \phi^2 \partial_\lambda b_\mu, \tilde g^{0\nu\prime} \phi^{2\prime} B^\prime ] \Bigr)
\nonumber\\
&{}& \equiv 2 ( I + J ).
\label{Direc-method}  
\end{eqnarray} 
It is straightforward to compute $J$ whose final result takes the form
\begin{eqnarray}
J = i \int d^3 x \, \delta_\mu^0 \tilde g^{\lambda\nu} \phi^2 \partial_\lambda B.
\label{J-result}  
\end{eqnarray} 
Here, as well as the extended de Donder gauge condition (\ref{Ext-de-Donder}), we have used the following ETCRs:
\begin{eqnarray}
&{}& [ \dot b_\mu, \tilde g^{0\nu\prime} ] = - i \Bigl\{ [ \tilde f \phi^{-2} \partial_\mu \tilde g^{0\nu}
+ \partial_0 ( \tilde f \phi^{-2} ) \tilde g^{00} \delta_\mu^\nu ] \delta^3
+ ( \tilde g^{0k\prime} \delta_\mu^\nu - \tilde g^{\nu k\prime} \delta_\mu^0 
+ \tilde g^{0\nu\prime} \delta_\mu^k ) \partial_k^\prime ( \tilde f \phi^{-2} \delta^3 ) \Bigr\},
\nonumber\\
&{}& [ \dot b_\mu, \phi^{2\prime} ] = - 2 i \tilde f \phi^{-1} \partial_\mu \phi \delta^3,
\qquad
[ \dot b_\mu, B^\prime ] = - i \tilde f \phi^{-2} \partial_\mu B \delta^3. 
\label{J-ETCR}  
\end{eqnarray} 
  
Next, let us pay our attention to an evaluation of $I$, which can be divided into four parts:
\begin{eqnarray}
&{}& I = \int d^3 x d^3 x^\prime x^{\nu\prime} \Bigl( [ \tilde g^{00} \phi^2 \dot b_\mu,
\tilde g^{00\prime} \phi^{2\prime} \dot B^\prime ]
+ [ \tilde g^{0i} \phi^2 \partial_i b_\mu, \tilde g^{00\prime} \phi^{2\prime} \dot B^\prime ] 
\nonumber\\
&{}& + [ \tilde g^{00} \phi^2 \dot b_\mu, \tilde g^{0j\prime} \phi^{2\prime} \partial_j B^\prime ]
+ [ \tilde g^{0i} \phi^2 \partial_i b_\mu, \tilde g^{0j\prime} \phi^{2\prime} \partial_j B^\prime ] \Bigr)
\nonumber\\
&{}& \equiv I_1 + I_2 + I_3 + I_4.
\label{I-4-part}  
\end{eqnarray} 
The calculation of $I_4$ is easiest since it involves only one ETCR as follows:
\begin{eqnarray}
I_4 &=& \int d^3 x d^3 x^\prime x^{\nu\prime} \tilde g^{0i} \phi^2 \phi^{2\prime} \partial_j B^\prime 
[ \partial_i b_\mu, \tilde g^{0j\prime} ]
\nonumber\\
&=& - i \int d^3 x \delta_\mu^i x^\nu \partial_0 ( \tilde g^{00} \phi^2 ) \partial_i B.
\label{I_4}  
\end{eqnarray} 
In a similar manner, it is straightforward to carry out the calculations of $I_2$ and $I_3$,
which are summarized as
\begin{eqnarray}
I_2 &=&  i \int d^3 x \tilde g^{0i} \phi^2 \delta_\mu^j ( \delta_i^\nu \partial_j B 
+ x^\nu \partial_i \partial_j B ),
\nonumber\\
I_3 &=&  i \int d^3 x \Bigl[ x^\nu \phi^2 ( \tilde g^{00} \ddot B \delta_\mu^0 
+ \tilde g^{0i} \partial_i \partial_\mu B ) - \delta_i^\nu \phi^2 ( - \tilde g^{0i} \delta_\mu^j 
\nonumber\\
&+& \tilde g^{ij} \delta_\mu^0 - \tilde g^{0j} \delta_\mu^i ) \partial_j B \Bigr].
\label{I_23}  
\end{eqnarray} 

The evaluation of $I_1$ is more difficult than those of the other $I_i ( i = 2, 3, 4 )$, but
is straightforward whose result is given by
\begin{eqnarray}
I_1 = - i \int d^3 x x^\nu \Bigl[ \partial_\mu ( \tilde g^{00} \phi^2 \dot B ) 
+ \delta_\mu^i \partial_0 ( \tilde g^{00} \phi^2 ) \partial_i B \Bigr],
\label{I_1}  
\end{eqnarray} 
where in particular we have used the ETCR: 
\begin{eqnarray}
[ \dot b_\mu, \dot B^\prime ] = - i \partial_0 ( \tilde f \phi^{-2} \partial_\mu B ) \delta^3 
- 2 i \tilde f \tilde g^{0i} \partial_i ( \tilde f \phi^{-2} \partial_\mu B \delta^3 ).
\label{db-dB}  
\end{eqnarray} 
Then, summing up $I$ and $J$, and multiplying it by 2, we can obtain
\begin{eqnarray}
[ P_\mu, K^\nu ] &\equiv& 2 ( I + J )
\nonumber\\
&=& 2 i \int d^3 x \, \tilde g^{0\lambda} \phi^2 \partial_\lambda B \delta_\mu^\nu
\nonumber\\
&=& - 2 i D_0 \delta_\mu^\nu 
\nonumber\\
&=& - 2 i ( G^\rho{}_\rho - D ) \delta_\mu^\nu.
\label{Direc-method2}  
\end{eqnarray} 


\end{document}